\documentclass[preprint]{aastex631}

\usepackage{bm}
\usepackage{amsmath}
\usepackage{multirow}
\usepackage{url} 
\usepackage[inline]{trackchanges} 
\usepackage{soul}

\newcommand{\beq}{\begin{equation}}
\newcommand{\eeq}{\end{equation}}

\def\la{\hbox{\raise.35ex\rlap{$<$}\lower.6ex\hbox{$\sim$}\ }}
\def\ga{\hbox{\raise.35ex\rlap{$>$}\lower.6ex\hbox{$\sim$}\ }}

\def\water{H$_2$O }

\def\Keff{K_{{\rm{eff}}}}
\def\Pvap{P_{{\rm vap}}}

\def\beq{\begin{equation}}
\def\eeq{\end{equation}}
\def\beqa{\begin{eqnarray}}
\def\eeqa{\end{eqnarray}}

\def\order#1{{\cal O}\left({#1}\right)}

\newcommand{\REV}[1]{{#1}}

\newcommand{\REVNEW}[1]{{#1}}
\newcommand{\REVNEWW}[1]{{#1}}

\begin{document}

\title{Retention of CO Ice and Gas Within 486958 Arrokoth}

\author[0000-0002-4578-1694]{Samuel P.D. Birch}
\affiliation{Department of Earth, Atmospheric and Planetary Sciences, Massachusetts Institute of Technology, Cambridge MA, USA}
\affiliation{Department of Earth, Environmental and Planetary Sciences, Brown University, Providence RI, USA}

\author[0000-0001-5372-4254]{Orkan M. Umurhan}
\affiliation{SETI Institute, Mountain View CA, USA}
\affiliation{Space Sciences Division, Planetary Systems Branch, NASA Ames Research Center, Moffett Field CA, USA}
\affiliation{Cornell University, Ithaca NY, USA}
\affiliation{Earth and Planetary Sciences, University of California, Berkeley, Berkeley, CA, USA}

\correspondingauthor{Sam Birch (sambirch@brown.edu) \& Orkan Umurhan (oumurhan@seti.org)}

\defcitealias{NIST}{NIST}

\begin{abstract}
Kuiper Belt Objects (KBOs) represent some of the most ancient remnants of our solar system, having evaded significant thermal or evolutionary processing. This makes them important targets for exploration as they offer a unique opportunity to scrutinize materials that are remnants of the epoch of planet formation. Moreover, with recent and upcoming observations of KBOs, there is a growing interest in understanding the extent to which these objects can preserve their most primitive, hypervolatile ices. Here, we present a theoretical framework that revisits this issue for small, cold classical KBOs like Arrokoth. Our analytical approach is consistent with prior studies but assumes an extreme cold end-member thermophysical regime for Arrokoth, enabling us to capture the essential physics without computationally expensive simulations. Under reasonable assumptions for interior temperatures, thermal conductivities, and permeabilities, we demonstrate that Arrokoth can retain its original CO stock for Gyrs if it was assembled long after the decay of radionuclides. The sublimation of CO ice generates an effective CO `atmosphere' within Arrokoth's porous matrix, which remains in near vapor-pressure equilibrium with the ice layer just below, thereby limiting CO loss. According to our findings, Arrokoth expels no more than $\approx 10^{22}$ particles s$^{-1}$, in agreement with upper limits inferred from \textit{New Horizons}' 2019 flyby observations. While our framework challenges recent predictions, it can serve as a benchmark for existing numerical models and be applied to future KBO observations from next-generation telescopes.
\end{abstract}


\section{Introduction}
Comets and Kuiper Belt Objects (KBOs) are a diverse population of small icy bodies that contain varying amounts of primitive refractory and volatile materials within their interiors. This diversity is a result of the range of temperature environments that they inhabit, from the inner solar system where most ices sublimate quickly, to the outer solar system where most ices can remain frozen since the planet formation era.

\begin{figure}[h!]
\noindent\includegraphics[width=\textwidth]{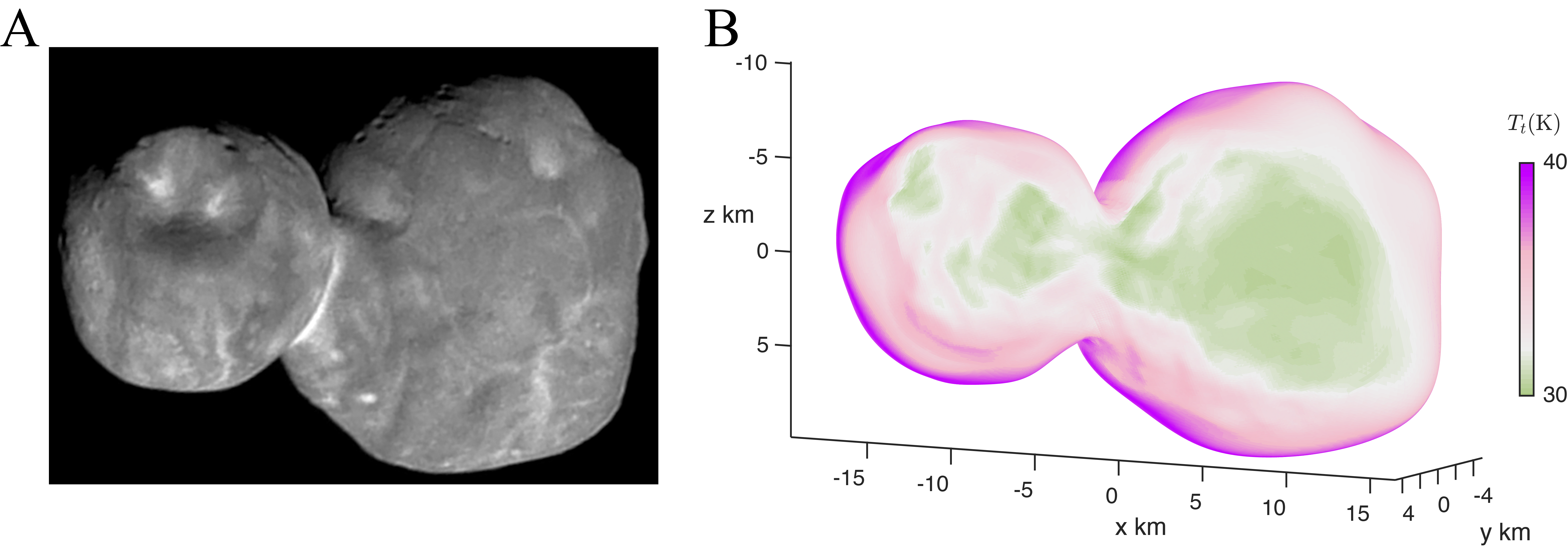}
\caption{A: Captured by the Multicolor Visible Imaging Camera (MVIC) component of the Ralph instrument onboard \textit{New Horizons}, this image was taken on January 1, 2019, 7 minutes prior to the spacecraft's closest approach, which was at a distance of 6700 kilometers from the surface. Credit: NASA/Johns Hopkins University Applied Physics Laboratory/Southwest Research Institute; B: Orbitally averaged temperature at the seasonal skin depth $r_t$, which was computed according to the approach detailed in \cite{umurhan_near-surface_2022}.  The physical scale of Arrokoth is shown in kilometers, while the orientation is comparable to that of panel A, looking down on the south pole.}
\label{arrokoth_figure}
\end{figure}

As members of the family of cold classical KBOs, 486958 Arrokoth (hereafter Arrokoth) was observed by the \textit{New Horizons} spacecraft \citep[Figure \ref{arrokoth_figure}A;][]{Stern_etal_2019} and provides a unique window into the earliest stages of the solar system. Arrokoth is one of the most primitive objects in our solar system, having never been significantly heated within the inner solar system. Its bi-lobate structure, composition, shape, and dynamical family suggest that it has likely remained in its current orbit since its formation \citep{spencer_geology_2020, grundy_color_2020,keane_geophysical_2022,mckinnon_solar_2020}. Arrokoth may have formed well into the solar nebula's Class II epoch, more than 4 million years after CAIs \citep{bierson_using_2019}. Its chemistry may therefore most closely resemble that of the protostellar and protoplanetary disk environment from which it formed, which is rich in hypervolatile ices like CO \citep{chiar1994,pontoppidan_spatial_2006,caselli_co_1999,mcclure_ice_2023}.

Many numerical models have been developed to understand the activity of small bodies like comets and centaurs and how various ices and gases may be retained \citep[][to name only a few]{Prialnik_etal_2004,gkotsinas_thermal_2022,jindal_topographically_2022,loveless_structure_2022,lisse_origin_2021,davidsson_thermophysical_2021,davidsson_collisional_2023,merk_combined_2006,steckloff_sublimative_2021,bouziani_cometary_2022,malamud_are_2022,Parhi_Prialnik_2023}. Such models are required because comets and centaurs especially can have multiple heat sources that drive the transport of sublimated vapor through a porous matrix that initially retains numerous ices with varying vapor pressures \citep{Prialnik_etal_2004}. This leads to a highly non-linear system. Sophisticated models like NIMBUS \citep{davidsson_modelling_2022} provide great detail, having been tested and calibrated on fine-resolution data acquired by \textit{Rosetta} at comet 67P/Churyumov-Gerasimenko. Often, these models show that comets (and centaurs) should be strongly active at large heliocentric distances (20$-$30 AU) if they retained significant hypervolatile ices. However, despite CO production potentially outpacing CO$_2$ in some objects \citep[e.g., 1P/Halley,][]{woods_rocket_1986}, particularly at greater heliocentric distances \citep{mumma_chemical_2011,womack_co_2017,chandler_cometary_2020,pinto_survey_2022}, the vast majority of small bodies \citep[outside objects like C/2016 R2 (PanSTARRS) and 29P/Schwassmann–Wachmann,][]{roth_molecular_2023, cordiner_sublime_2022} do not show substantially elevated CO activity levels. Primordial CO in these more close in objects may therefore be largely lost \citep[e.g.,][]{Parhi_Prialnik_2023}, with modern day CO released due to the sublimation of entrapping, less volatile ices, possibly as nm-scale intimate mixture within amorphous \water or CO$_2$ or photodissociation of other more abundant ices like CO$_2$ \citep[e.g.,][]{rubin_origin_2020,luspay-kuti_dual_2022}.

KBOs, like comets and centaurs, are highly porous bodies \citep{keane_geophysical_2022} and likely have both volatile and refractory materials intimately mixed within their interiors. However, KBOs may even retain their most primitive volatiles, as they have not experienced significant solar heating at large heliocentric distances. Observations of old debris disks (e.g., in Fomalhaut and HD181327) indicate the presence of non-primordial (i.e., secondarily sourced) CO gas, interpreted as possibly originating from extrasolar Kuiper belt and/or exocometery objects \citep{Matra_etal_2017,Kral_etal_2017, Wyatt_2020}.  
Theoretical modeling of grain growth during the planetesimal formation phase of our solar system strongly suggests CO as a major constituents of all planetesimals formed at these distant locations \citep[e.g.,][and references therein]{Estrada_Cuzzi_2022}, and that planetesimals are ideal locations to sequester such volatiles \citep[e.g.,][]{Krijt_etal_2020}.
Upcoming JWST observations of KBOs of our own solar system will soon reveal the existence or absence of CO and other volatiles associated with these bodies. 

It is therefore surprising that CO was not detected within the limits of the \textit{New Horizons} spacecraft \citep{Stern_etal_2019}, leading to suggestions that CO may be depleted in even distant, cold small bodies, prior to their injection into the inner solar system \citep{lisse_origin_2021,Parhi_Prialnik_2023}. However, we must consider that \textit{New Horizons} provided only coarse lower limits, which could allow for scenarios where Arrokoth may be weakly outgassing enough CO that could be detected with more sensitive instrumentation, as recently suggested by the calculations presented in \cite{Kral_etal_2021}. Another confounding observation is that methanol ice was also detected on Arrokoth's surface, but no evidence of water ice was found \citep{grundy_color_2020}.
\REV{Is CO involved?} 
Taken together, these puzzling observations make it unclear to what extent the most volatile ices have been retained within KBOs, how various thermal evolutionary processes in the distant solar system since their formation may have depleted their initial inventory, and how representative modern outgassing measurements are of their bulk ice inventories.

To address these questions, we have crafted an analytical framework that is in harmony with extensive previous research (Section \ref{Section2}), operating within an extreme end-member scenario where Arrokoth is incredibly frigid \citep[\REV{$<$40K}, see Figure \ref{arrokoth_figure}B;][]{umurhan_near-surface_2022}. Instead of relying on computationally intensive models, we have opted for a simplistic analytic framework that captures the essential physics, whereby the numerous non-linear feedbacks and other short timescale physics are of lesser importance (Section \ref{timescales_analysis}). We have subsequently made estimations concerning the viability of CO ice within Arrokoth $-$ on the assumption that certain planetesimals of the cold classical Kuiper Belt, like Arrokoth, were formed after the decay of radionuclides $-$ and have demonstrated that significant volumes of its original CO ice and gas can still be retained within its interior up to the present day (Section \ref{Section3}). Our methodology showcases that Arrokoth, and conceivably other KBOs, retain a CO `atmosphere' within their porous interiors, with only weak outgassing that falls below current detection thresholds. Furthermore, the solutions obtained from our analytical framework can subsequently be used to verify and benchmark more complex models (Section \ref{Section4}), and can be expanded to future KBO observations conducted by next-generation space- and ground-based telescopes.

\section{Analytical Model for Sublimation-Driven Gas Transport in Arrokoth-Like Objects} \label{Section2} 
\begin{figure}[h!]
\noindent\includegraphics[width=\textwidth]{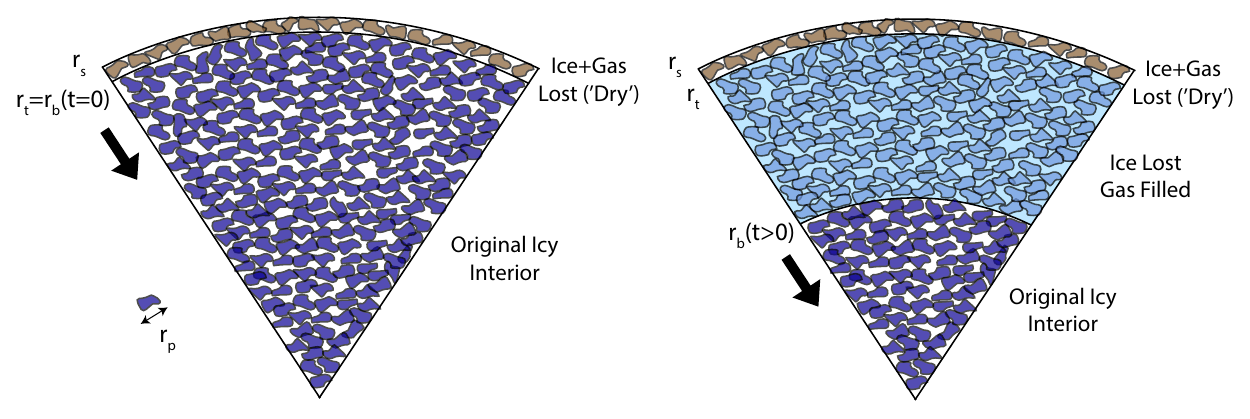}
\caption{Model setup: A porous rubble pile, with the solid matrix comprised of intimately mixed CO ice and refractory amorphous \water ice, with pore radii $r_p$. The top-most layer (brown) is thermally processed within a single orbit, with any volume of the assumed CO (ice and gas) absent. Material below the sublimation front $r_b$ (dark blue) retains its initial CO ice volume. As the sublimation front migrates downward through time (right), CO within the the amorphous \water ice matrix sublimates. The produced gas (light blue) fills the pore space and migrates radially upward, away from the sublimation front.}
\label{model_cartoon}
\end{figure}

We are considering a spherical body with a radius of $r=r_s$, and we assume that the seasonal thermal skin depth is located at $r=r_t$ (see Figure \ref{model_cartoon}). At this location, we take the temperature to be the average of the extreme high ($T=T_{{\rm max}}=58K$) and low ($T=T_{{\rm min}}=12K$) temperatures of the surface, which is given by $T_t\equiv T(r=r_t)=0.5(T_{{\rm max}}+T_{{\rm min}})$. We keep this temperature fixed, and use it as the initial upper boundary condition for modeling the evolution of Arrokoth's bulk interior. Based on our spherical approximation of Arrokoth's shape and fixed orbit, we can assert that this simplifying assumption is valid \citep[see Figure \ref{arrokoth_figure}B;][]{umurhan_near-surface_2022}. All variables henceforth are listed in Table \ref{values_table}. 

Assuming that Arrokoth's structure below $r_t$ is a rubble pile with a global porosity of $\Psi$, we consider the solid portion of the body as composed of non-volatile weakly self-adhering material (amorphous \water ice) and a single volatile ice species (CO), which constitutes a fraction $f_i$ of the total mass density. \REV{More complex interior structures can be considered in follow-up work.} Our model assumes a thin impermeable sheen at $r=r_b$ for the volatile-ice refractory matrix, from which the ice sublimates steadily. This is a slight underestimation of the sublimating surface area compared to other models \citep{davidsson_thermophysical_2021}. At $t=0$, $r_b=r_t$. As the ice sublimates, the sublimation front moves deeper into the interior, and the gas slowly flows upward through an ice-free porous matrix above $r_b$, characterized by a pore size of $r_p$. Notably, the material below $r_b$ always retains its initial CO ice inventory, as illustrated in Figure \ref{model_cartoon}. 

We postulate that the CO gas flow through the porous matrix is driven by pressure gradients that arise from the sublimation process at the front $r_b$ (as depicted in Figure \ref{model_cartoon}). Gas between $r_t$ and $r_b$ can be stored for long timescales, and only slowly leaks out. All gas above $r_t$ is assumed to be rapidly lost. As the sublimation front $r_b$ migrates downwards, its rate slows progressively as a decreasing amount of energy is conducted downwards. To simplify our analysis, we assume that the solid matrix above $r_b$ is unaffected by the sublimation of ices, so the pore sizes remain fixed through time. Although processes such as sublimation/re-condensation, hydrostatic closure with increasing depth, fracturing \citep{el-maarry_fractures_2015}, or removal of overburden material via large-scale collapse \citep[i.e., sinkholes;][]{vincent_large_2015} can affect the grain pore sizes, we expect the strength of analogous cometary-like material \citep{groussin_thermal_2019}, which we assume is similar for Arrokoth, to render such processes unimportant \REV{for our initial work here} (see Section \ref{sec:appendix}). The dynamic evolution of pore radii may be an outcome of the flow of gases through the interior for warmer or smaller icy small bodies  
\citep[e.g.,][]{Parhi_Prialnik_2023}, which we leave as a topic for future studies.

{\it{Key to our subsequent analyses}}: we assume that the evolution of $r_b$ is a quasi-static process, where $r_b/\dot r_b$ progresses \REVNEW{on front propagation timescales ($\tau_{fr}$)} much longer than both the thermal ($\tau_t$) and dynamical ($\tau_d$) readjustment timescales for the region below $r=r_t$ and above $r=r_b$. In Section \ref{timescales_analysis}, we will explore the conditions under which this assumption may fail. Nevertheless, we confirm that this condition is met for an object like Arrokoth (Section \ref{timescales_analysis}), allowing us to use our simple analytic treatment instead of more computationally expensive numerical models.

Thus, not unlike in previous work \citep[e.g.,][]{davidsson_thermophysical_2021,davidsson_modelling_2022,bouziani_cometary_2022,Parhi_Prialnik_2023}, the long-term migration rate of the sublimation front $r_b$ is controlled by three fundamental concepts: downward heat conduction, upward gas flow, and the conservation of mass. 

\subsection{Thermal evolution}
Describing the thermal evolution of the interior of Arrokoth, we use
\beq
0 = \frac{1}{r^2}\frac{\partial}{\partial r} 
K_{{\rm eff}} r^2 \frac{\partial T}{\partial r}
-  \dot {{\cal Q}} \delta(r-r_b),
\label{thermalconduction}
\eeq
where $r$ denotes the radial distance from the core ($r=0$) to the surface ($r=r_s$) of Arrokoth (refer to Figure \ref{model_cartoon}). The temperature $T$ depends on $r$ and is determined by the effective conductivity ($\Keff$), which in general includes heat transfer through the solid matrix and radiative conduction across pore spaces, the latter being negligible for the cold conditions we consider here. We also neglect heat transfer through the gas phase (see Section \ref{sec:appendix}). Several previous theoretical, observational, and experimental studies \citep{groussin_thermal_2019,umurhan_near-surface_2022,bouziani_cometary_2022,davidsson_modelling_2022,Parhi_Prialnik_2023} provide a range of plausible effective conductivities. 

We assume that the solution to the heat equation is constrained by a fixed temperature, $T_t$, at the location of the seasonal thermal skin depth. The heat loss term, $\dot Q$ (Eq. \ref{heat_equation}), is represented by the Dirac delta function at $r=r_b$, with the coefficient $\dot{\cal Q}$ quantifying the energy consumed in driving sublimation. This is formally treated as a boundary condition at $r=r_b$, i.e.,
\beq
\Keff\partial_r T\big|_{r=r_b} = \dot {{\cal Q}} = {\cal L} \dot\Sigma(r=r_b),
\label{Tb_bc}
\eeq
where ${\cal L}$ represents the enthalpy of sublimation per unit mass of CO ice, and $\dot\Sigma$ is the net mass loss rate per unit area (specifically, in units of kg$\cdot$m$^{-2}\cdot$s$^{-1}$). To model $\dot{\Sigma}$ at $r=r_b$, we use a modified version of the method proposed by \cite{lebofsky_stability_1975}, where gas production is proportional to the difference in ambient gas pressure just above $r=r_b$ and the vapor pressure generated by ice sublimation at $r=r_b$.  In other words
\beq
\dot\Sigma = \frac{v_k}{{c_s}^2} \Delta P = \frac{v_k}{{c_s}^2}\Bigg( P_{vap}\Big(T_b\Big) - P_b \Bigg),
\label{lebovsky_label}
\eeq
where the sublimated gas molecules have a mean-averaged kinetic speed $v_k \equiv \sqrt{8/\pi} c_s$, \REVNEW{$c_s^2 \equiv P/\rho$ is the square of the isothermal sound speed}, $R_g = 8315$ J/kg/K is the gas constant, and $\mu_a$ is the average atomic weight of the gas molecules. $P_b$ represents the ambient gas pressure at $r=r_b$, and $P_{vap}(T_b)$ is the saturation vapor pressure at temperature $T_b \equiv T(r=r_b)$.  
\REV{$\dot\Sigma$ is generally expressed as the product of $v_k$ and the net sublimating gas density, $\Delta\rho$, the latter of which has been re-expressed in Eq. (\ref{lebovsky_label}) in terms of $c_s^2$ according to the ideal gas law, i.e., $\Delta P = c_s^2\Delta\rho$.}

We further make the assumption that there are no internal heat sources, such as from the radiogenic decay of $^{26}{\rm Al}$ or the exothermic transition of amorphous to crystalline water ice. This assumption is reasonable for cold-classical KBOs like Arrokoth (refer to Section \ref{Section3}) because they may have formed late from the protoplanetary disk \citep[e.g.,][]{bierson_using_2019} and may have never been heated above $\sim 60$ K \citep{umurhan_near-surface_2022}. \REV{As our work is focused on building the initial framework for the evolution of CO within cold KBOs, the potentially non-negligible impact of both short- and long-lived radionuclides \citep[e.g.,][]{davidsson_thermophysical_2021} will be treated in follow-up numerical studies.}

We use Eq. (\ref{thermalconduction}) under the implicit assumption that $\rho_{tot} C_p(dT/dt)=0$, where $\rho_{tot}$ represents Arrokoth's bulk density and $C_p$ is the specific heat at constant pressure of the solid ice constituent of the CO-depleted layer above $r_b$. This assumption indicates that Arrokoth attains a thermal steady-state from its initial cold formation temperature much faster than it sublimates all of its CO, which results from being in the extreme $\tau_t\ll\tau_{fr}$ limit. Although numerical modeling that retains the $\rho_{tot} C_p(dT/dt)$ term always produces correct thermal profiles, any errors incurred by its neglect -- as we have done here --  will be insignificant for very cold bodies such as Arrokoth, where the $\tau_t\ll\tau_{fr}$ condition is met. This is elaborated in Section \ref{timescales_analysis}.

It is worth reiterating that the boundary condition in Eq. (\ref{Tb_bc}) specifies that all the thermal flux reaching the sublimating front is completely used up there. In more general treatments that include the $\rho_{tot} C_p(dT/dt)$ term in the heat equation, it is necessary to consider the difference in thermal fluxes on both sides of the front to accurately account for the energy that drives sublimation. Our approach to the thermal boundary condition may seem to provide an excess of this energy, but this is not the case because our method assumes that $\tau_t \ll \tau_{fr}$, which means that the system has had enough time to reach its quasi-static state. In \ref{sec:appendix}, we demonstrate that in this quasi-steady state, the temperature configuration below the front is simply $T(r<r_b) = T_b$, which, in turn, implies that the thermal flux immediately beneath the front is zero. Therefore, in this extreme timescale limit, all the incoming thermal flux from above the front is consumed to drive sublimation there. The reason for this is apparent: there is enough time to raise the temperature of the interior to its natural spatially uniform value $T_b$ before there is any significant movement in the position of $r_b$ caused by sublimation.

\subsection{Gas flow through a porous medium}
The Darcy flow law, corrected for Knudsen diffusion \citep{ziarani_knudsens_2012}, governs the spherically symmetric gas flow through the porous matrix as
\beq
0 = -{k_a}\frac{\partial P}{\partial r}  -{\mu} u,
\label{darcy}
\eeq
where a pressure gradient $\frac{\partial P}{\partial r}$ drives the flow of gas of viscosity $\mu$ and radial velocity $u$ through a matrix with an effective permeability $k_a$. We disregard gravitational effects in our formulation since we observe that $\rho_{tot} g$ is generally much smaller than $\partial P/\partial r$ for small bodies with sizes less than $100$ km. 

The formula given by \citep{ziarani_knudsens_2012} modifies the effective permeability based on the empirically-derived Knudsen number
\beq
k_a = k_{\infty}(1+4 \bar c {\rm Kn}),
\label{knudsen}
\eeq
where $\bar c$ is an $\order{1}$ constant, and ${\rm Kn} = \lambda / r_p$ is the Knudsen number defined in terms of the mean free path of vapor molecules, $\lambda = 1/n\sigma = {m}\big/{\rho \sigma}$, with $\sigma$ being the cross-section of collision of gas molecules with number density $n$. At very small Knudsen numbers, gas diffusion through pore spaces is dominated by molecule-molecule collisions, and the effective `liquid' permeability reduces to the Darcy limit $k_{\infty} = {r_p}^2/32$ \citep{bouziani_cometary_2022}. However, for larger Knudsen numbers, where the mean free path is comparable to or larger than the pore radius, gas diffusion is instead dominated by molecule-matrix collisions. \REVNEWW{In our work, we are often in the large Knudsen number limit.} 

Equation (\ref{darcy}) requires a boundary condition, for which we assume that the ambient gas pressure is zero at the surface, i.e., $P(r=r_s) = 0$. We also assume that the interior gas pressures ($\approx {10^{-4}}$ Pa) are weaker than the bonding strengths ($\sigma_b$) between individual $\mu$m-scale grains
\citep[$\approx 1$ kPa,][]{Gundlach_etal_2018}
and/or their mm-scale aggregates 
\citep[$\approx 1$ Pa,][]{Blum_etal_2014}. This justifies the use of Eq. (\ref{darcy}). For further details, see \ref{sec:appendix}. 

Our approach for the Kn$\gg 1$ regime is analogous to the Skorov-Rickman formulation 
\cite{Skorov_Rickman_1995} for high-Kn flow, which can be found in Eqs. (2) \& (46) of \cite{davidsson_thermophysical_2021}. Generally, the Skorov-Rickman formulation is dependent on pores that are tubes with length ($L_p$), width ($r_p$), and tortuosity ($\xi$). 
\REVNEW{Due to a lack of non-theoretical constraints on $L_p$ in putatively pristine planetesimals like Arrokoth,
for} the present investigation we adopt the assumption that $\xi = 1$ and $L_p = r_p$.
\footnote{\REVNEW{According to \cite{Skorov_Rickman_1995} the diffusion coefficient $D\sim (2/3)r_p$ for $L_p = r_p$, while for $L_p\gg r_p$ it follows that $D\sim (8/3)r_p$, meaning that in the latter limit, the diffusion rate would be enhanced by a factor of $4\times$, thereby reducing the volatile lifetime by that factor.}}
Further investigation of the overall dependencies on these parameters, utilizing our quasi-static model, remains a task for future studies.

\subsection{Mass conservation}
Finally, mass conservation within the object is maintained using the equation
\beq
\frac{1}{r^2}\partial_r r^2 \rho u = \dot{\Sigma} \delta(r-r_b),
\label{massconservation}
\eeq
where $\dot{\Sigma}$ is the instantaneous mass loss rate defined in Eq. (\ref{lebovsky_label}), and $\rho$ is the gas density. The solution of Eq. (\ref{massconservation}) yields
\beq
\rho u = \frac{{r_b}^2}{r^2} \frac{v_k}{{c_s}^2}\Bigg( P_{vap}\Big(T_b\Big) - P_b \Bigg),
\label{lebovsky}
\eeq
which, together with Eq. (\ref{darcy}), establishes a solution for $P$ and subsequently $P_b$ (see details in Section \ref{sec:appendix}).

Lastly, the rate at which the sublimation front propagates through the volatile ice portion of the total matrix is specified as
\beq
\rho_{ice} \dot r_b = \frac{v_k}{{c_s}^2}\Bigg( P_{vap}\Big(T_b\Big) - P_b \Bigg),
\label{dotrb_formal}
\eeq
where $\dot r_b \equiv dr_b\big/dt$. We account for the reduced porosity and dust-to-ice fraction in the partial ice density $\rho_{ice}$.

Net sublimation will occur when $P_b < P_{vap}(T_b)$. For an Arrokoth-sized object with low internal temperatures \citep[Figure \ref{arrokoth_figure}B;][]{umurhan_near-surface_2022}, $P_{vap}(T_b) - P_b$ is typically $\order{10^{-5}}$ Pa or smaller. Significant sublimation requires long timescales for these small deviations from the saturation vapor pressure, which may pose challenges for efficient time-stepping in numerical models that include the time derivative in the thermal energy equation. This highlights the usefulness of our analytic approach (Section \ref{Section4}). 

Our approach differs from recent works such as \cite{bouziani_cometary_2022} and \cite{lisse_origin_2021} (see \ref{app:2_lisse}), who assumed $P_b = 0$ in Eq. (\ref{lebovsky}). However, our formulation is broadly consistent with standard thermophysical models  \citep[e.g.,][among many others]{Prialnik_etal_2004,davidsson_thermophysical_2021} that have built on decades of prior research.

\subsection{Implementation}
Combining Eqs. (\ref{thermalconduction}$-$\ref{dotrb_formal}), and imposing $P=0$ at $r=r_s$, yields a single ordinary differential equation for the time-dependent rate at which $r_b$ propagates into the interior. \REVNEW{The methodology for solving this equation is algebraicly involved and, as such, its details are relegated to \ref{sec:appendix}.} The temperature solution across the entire domain is given by Eqs. (\ref{Tr_solution}-\ref{Tr_below_solution}) and Eq. (\ref{Fb_solution}). The resulting approximate differential equation describing the sublimation front's propagation rate, normalized by Arrokoth's radius, $\zeta$, follows as
\beq
\zeta (1 - \zeta) \dot{\zeta} = \frac{1}{6 \tau_{s}(T_b)},
\qquad \zeta \equiv \frac{{r_b(t)}}{r_s},
\label{rate}
\eeq
where 
\beq
\tau_s(T_b,r_p,r_s) \equiv  \frac{\REVNEWW{3}{r_s}^2  \rho_{{\rm ice}} v_k}{\REVNEWW{8} r_p \bar{c} \Pvap(T_b)}\Bigg/
\left[{\displaystyle{1 + \frac{\Pvap(T_b)}{\tilde P} }}
\right],
\label{timescale}
\eeq
is the timescale of the volatile ice layer's total lifetime, and a transition pressure
\beq
\tilde{P} = \frac{8m {c_s}^2 \bar{c}}{\sigma r_p} = \frac{24 c_s^2 \bar c}{v_k r_p}\mu,
\label{Ptilde}
\eeq
\REV{delineating the switch from fluid flow (Kn $< 1$) to diffusive molecular flow (Kn $\gg 1$).}
\REVNEW{This timescale results from the combination of Eqs. (\ref{dotSigma}-\ref{dotrb}), which constitutes the final step in the derivation of the governing ordinary differential equation.}
When writing Eqs. (\ref{timescale}-\ref{Ptilde}), we explicitly used the Maxwell kinetic theory approximation for CO gas viscosity, where $\mu \approx m v_k/3\sigma$. The timescale $\tau_s$ depends on $T_b$,  which is a diagnostic function of $\zeta$ and $T_t$, given by 
\beq
\displaystyle
T_t-T_b = 
\overline{\Delta T}
\left[\frac{\Pvap(T_b)}{\Pvap(T_t)}\right]
\left[1 + \frac{\Pvap(T_b)}{\tilde P} \right]
\left( \frac{1-\zeta \psi^{-1}}{1-\zeta}\right)
,
\label{DeltaT_diagnostic}
\eeq
with
\beq
\displaystyle{\overline{\Delta T}} \equiv
\frac{3r_p \bar c {\cal L}}{8v_k\Keff} \Pvap\left(T_t\right).
\qquad
\psi \equiv r_t/r_s.
\label{DeltaT_diagnostic_and}
\eeq
We use the characteristic difference temperature scaling ${\overline{\Delta T}}$ at the top of the volatile ice layer and introduce the ratio $\psi$ ($<1$) representing the base of the seasonal thermal wave penetration depth to the total radius of Arrokoth in Eqs. (\ref{timescale}-\ref{Ptilde}). Although there is some dependence on $\zeta$ near $\zeta=\psi$ in Eq. (\ref{DeltaT_diagnostic}), it is negligible elsewhere. Henceforth, we assume $({1-\zeta \psi^{-1}})\big/({1-\zeta}) \rightarrow 1$, which makes it easier to derive an analytical solution since $\tau_s(T_b)$ becomes independent of $\zeta$. As a result, Eq. (\ref{rate}) directly integrates into an implicit relationship for $\zeta$, where we take $\zeta(t=0) = 1$
\beq
3\zeta^2 - 2\zeta^3 = 1 - \frac{t}{\tau_s(T_b)}.
\label{implicit_zeta_solution}
\eeq

When utilizing the analytical solutions provided in Eqs. (\ref{rate}-\ref{implicit_zeta_solution}), it is straightforward to apply them if the following condition is met
\beq
P_{{\rm lim}} \equiv 
\frac{2 r_s v_k \mu}{k_{\infty}}\zeta(1-\zeta) = \frac{64 r_s v_k \mu}{r_p^2} \zeta(1-\zeta)\gg
\left\{ 
\tilde P, \Pvap\left(T_t\right)
\right\},
\label{Plim_def_text}
\eeq
\REVNEW{where $P_{{\rm lim}}$ is a limiting pressure that emerges naturally from the solution procedure. Its physical meaning remains unclear.}
In our particular scenario, $P_{{\rm lim}} \ge \order{10^7 {\rm Pa}}$. This condition is usually satisfied when CO is retained within a KBO like Arrokoth (see \ref{sec:appendix}).

\section{Validity Criteria for Quasi-Static Evolution Theory} \label{timescales_analysis}
We assumed that the movement of gas and heat can be treated as quasi-static processes in response to the slow downward evolution of the CO ice front. This enabled us to obtain simple analytical solutions without relying on computationally intensive models. Although this assumption may not apply to all small, porous bodies and ices in the solar system, and numerical models will be needed to address such non-linear problems \citep[e.g., NIMBUS, as described in][]{davidsson_thermophysical_2021}, we believe it to be valid for cold bodies like Arrokoth. In the following section, we present an analysis of the conditions under which this assumption holds.

In order for the quasi-static theory discussed in Section \ref{Section2} to be valid, both $\tau_{fr} \gg \tau_t$ and $\tau_{fr} \gg \tau_d$ must hold. For a spherically symmetric object undergoing thermal diffusion, the slowest thermal relaxation time in a spherical shell bounded by $r_b$ and $r_t$ can be approximated by \citep[e.g.,][]{Coradini_etal_1997}
\beq
\tau_t \approx \frac{\rho_{tot} C_p}{\Keff}\frac{\Delta r^2}{\pi^2} = 
\frac{\rho_{tot} C_p}{\Keff}\frac{ r_s^2}{\pi^2}\left(1-\frac{r_b}{r_s}\right)^2
,
\label{thermal_timescale}
\eeq
where $\Delta r$ is defined as $r_t - r_b$ (which is approximately equal to $r_s-r_b$). For our calculations, we utilize the general formulation presented in \cite{Shulman_2004} for amorphous \water ice and take values within the temperature range $T_t$ as listed in Table \ref{values_table}. The thermal relaxation time $\tau_t$ typically ranges from $30$Kyr to $30$Myr, depending on various parameters such as $\Keff$ and $r_s$  \citep[e.g., see recent discussions in][and references therein]{davidsson_thermophysical_2021,davidsson_modelling_2022,Parhi_Prialnik_2023}. 

Likewise, in the case of gas density variations in a porous matrix with a pore size of $r_p$ and thermal velocities of $v_k$, the dynamical relaxation time takes the same form as before, but with a diffusion coefficient $D$ roughly equivalent to $\sim 3r_p v_k/8$
\citep[like the \cite{Skorov_Rickman_1995} formulation in][with $L_p = r_p$ and $\xi = 1$]{davidsson_thermophysical_2021}. 
Hence, we obtain
\beq
\tau_d \approx  \frac{1}{D}\frac{\Delta r^2}{\pi^2}.
\label{diffusive_timescale}
\eeq
For small objects, it is generally true that $\tau_d\ll\tau_t$ in the diffusive limit, thereby allowing the gas flow problem to be treated as a quasi-static process, as is done in this work and other studies  \citep[e.g.,][]{davidsson_thermophysical_2021}. Specifically, for pore sizes ranging from $0.01-1{\rm mm}$, thermal velocities of $v_k = \order{150 {\rm m/s}}$, and length scales of $\Delta r = \order{1{\rm km}}$, $\tau_d$ varies from $\approx10-1000$ years depending on assumed values of $r_p$.

We propose that $r_b/\dot r_b$ serves as a suitable approximate estimate for $\tau_{fr}$, \REVNEW{{\textit{as this is the gas loss timescale from the body}}}. 
By utilizing the relationship for $\dot \zeta$ from Eq. (\ref{rate}), we obtain
\beq
\tau_{fr} \equiv {\zeta}\Big/{\dot\zeta}  = 6\tau_s \zeta^2(1-\zeta).
\label{actual_sublimation_timescale}
\eeq
To establish the condition $\tau_{fr}\gg\tau_t$, we use the definition for $\tau_s$ from Eq. (\ref{timescale}) and perform some re-arrangement to obtain a requirement on $P_{{\rm vap}}(T_b)$ for the quasi-static evolution solutions to be valid. This condition is given by
\beq
P_{{\rm vap}}(T_b)\left(1+\frac{P_{{\rm vap}}(T_b)}{\tilde P}\right)
\ll
\frac{24\pi^2}{9\bar c}
\left(\frac{\rho_{ice}}{\rho_{tot}}\right)
\left(\frac{\Keff v_k}{r_pC_p}\right)
\frac{r_b^2\big /r_s^2}{1 - r_b/r_s}.
\label{general_validity_condition}
\eeq
The condition \REV{Kn $\gg 0.1$} is the same as $\tilde P \gg P_{{\rm vap}}(T_b)$, which implies that the quantity in the parenthesis on the left-hand side (LHS) of Eq. (\ref{general_validity_condition}) can be replaced by 1 in the Knudsen limit. This assumption is valid for objects such as Arrokoth, where the interior temperatures are low, and CO vapor pressures are small (Section \ref{Section3}). It is important to note that, for any value of $P_{{\rm vap}}(T_b)$, there exists a small enough $r_b/r_s<r_{b,min}/r_s$ where this approximation fails. However, for the conditions relevant to Arrokoth, we find that this breakdown only occurs for $r_b/r_s \le 0.05$ and $\tau_{fr}$ is $\order{10^8}$ years or more (see the next two Sections \& Figure \ref{validity_figure}A). 
\REV{We also call attention to the estimated sublimation timescales that go from being order of magnitude correct to being, instead, lower bounds whenever the timescale ordering transitions into $\tau_t\gg\tau_{fr}$, which may occur for conductivities as low as 10$^{-5}$ Wm$^{-1}$K$^{-1}$ as suggested for TNOs \citep{Lellouch_etal_2013}.}
\par
\REVNEW{The key physical feature is that the timescale of relevance is the front propagation timescale $\tau_{fr}$\REVNEWW{, which} is a consequence of the interior ice sublimation rate. This sublimation rate is {\textit{controlled}}
by the ambient gas pressure at the front.  It is for this reason that we do not consider the sublimation timescales often quoted based on the expression for $\tau_{{\rm{subl-H2O}}}$  found in Section 3.9 of \citet{Prialnik_etal_2004}, which estimates the time it takes the sublimating ice to equilibrate to the ambient pressure of the volatile gas.  Indeed, the gas loss timescale and subsequent volatile ice front propagation would be fast and given by $\tau_{{\rm subl-H2O}}$ if the ambient pressure were almost nearly zero (e.g., in a scenario where gas instantaneously streams out of the body). However, when the gas remains nearly trapped because it must move diffuse through a porous medium with small pore sizes, the loss timescale is governed by $\tau_{fr}$, \REVNEWW{which} is the relevant timescale for the dynamics considered here.}

\subsection{Validity at Arrokoth} \label{arrokoth_validity_section}
To confirm the validity of our quasi-static assumption for an object like Arrokoth, we can make use of established thermophysical properties to estimate important parameters. Firstly, we can estimate $\tilde P$ by inputting representative values relevant to our study, yielding
\beq
\tilde P \approx 7.8 \left(\frac{v_k}{165 \ {\rm m/s}}\right)
\left(\frac{1 \ {\rm mm}}{r_p}
\right) {\rm Pa},
\eeq
where the reference value for $v_k$ is chosen for the corresponding reference temperature of about 36K. Secondly, assuming that $\tilde P \gg P_{{\rm vap}}(T_b)$, we obtain a validity condition for pressures at $r_b$ from Eq. (\ref{general_validity_condition}), given by
\beq
P_{{\rm vap}}(T_b) \ll P_{{\rm val}} \equiv 
\displaystyle  \frac{r_b^2\big/r_s^2}{1-r_b\big/r_s} P_{\varphi},  
\label{small_tildeP_validity}
\eeq
\REVNEW{with $P_{\varphi}$ understood to be a reference limiting pressure derived from the solution validity condition
 $P_{{\rm vap}}(T_b)$ according to the ordering $\tau_s\gg\tau_{t}$, and is given by}
\beq
P_{\varphi}  \equiv 14.5\left(\frac{\rho_{ice}}{\rho_{tot}}\right)\left(\frac{1 \ {\rm mm}}{r_p} \right) 
\left(\frac{\Keff}{0.001 \ {\rm  W/m/K}}\right)
 \left(\frac{v_k}{165 \ {\rm m/s}}\right)
 \left(\frac{300 \ {\rm J/K/m}^3}{C_p} \right) 
\ {\rm Pa},
\eeq
and the specific heat value is chosen based on $C_p(T=36{\rm K})$ \citep{Shulman_2004}. \REVNEW{The pressure validity bound $P_{{\rm val}}$ is equal to $P_{\varphi}$ when $r_b/r_s = 1/\varphi = \varphi-1 \approx 0.618$, where $\varphi = (1+\sqrt 5)/2$ (the golden ratio).} Henceforth, we use $P_{\varphi}$ as a convenient reference for the pressure bound to ensure that the pressure at $r_b$ satisfies the validity condition for quasi-statically evolving solutions. 

We can define $r_{b,{{\rm min}}}$ as the depth at which $P_{vap}(T_b) = P_{{\rm val}}$. If $P_{vap}(T_b)\big/P_{\varphi} \ll 1$, then we have
\beq
\frac{r_{b,{{\rm min}}}}{r_s} \approx \sqrt{P_{vap}(T_b)\big/P_{\varphi}}.
\label{rb_min_definition}
\eeq
This implies that when $r_b \lessapprox r_{b,{{\rm min}}}$, the sublimation front is receding faster than the thermal relaxation time. The reason behind this can be attributed to the spherical geometry effect, where as the front gets closer to the core, the thermal energy conducted from $r=r_t$ increases as the inverse square, thus leading to an increase in the sublimation rate. Though our quasi-static evolutionary theory breaks down at a certain value of $r_b$, for Arrokoth, this breakdown occurs at very small values of $r_b/r_s$ (see Section \ref{Section3}).

\section{Estimating the Longevity and Outgassing Rates of CO at Arrokoth} \label{Section3}
Our hypothesis is that, although \textit{New Horizons} reported upper limits on outgassing rates during its flyby \citep[$\dot N< \dot N_{{\rm max}} = 3\times 10^{24}$ H atoms/s;][]{Stern_etal_2019}, \REVNEW{significant amounts of CO gas and ice may still exist within Arrokoth that can outgas below detection limits.} To investigate this hypothesis, we employ the system of equations outlined in Section \ref{Section2}, assuming that Arrokoth was initially seeded with substantial amounts of CO and that we remain in our quasi-static limit (Section \ref{timescales_analysis}). The input parameters are sourced from previous studies that analyzed \textit{New Horizons} data of Arrokoth and are summarized in Table \ref{values_table}.

We use conservative values for $r_s$ based on the best fit to Arrokoth's shortest dimension \citep[of both lobes Wenu and Weeyo;][Figure \ref{arrokoth_figure}A]{keane_geophysical_2022}. For the temperature range at the base of the seasonal thermal wave skin depth, we adopt temperature ranges quoted in \citep{grundy_color_2020,umurhan_near-surface_2022} and shown in Figure \ref{arrokoth_figure}B. The temperatures on the low end correspond to values found at Arrokoth's poles, which generally correspond to our above assumed values of $r_s$, while the higher temperatures correspond to the equatorial zones, where effective radii are larger 
\citep[10 km for Wenu and $7-8$ km for Weeyo;][see Figure \ref{arrokoth_figure}B]{keane_geophysical_2022}.

\begin{figure}[h!]
\noindent\includegraphics[width=\textwidth]{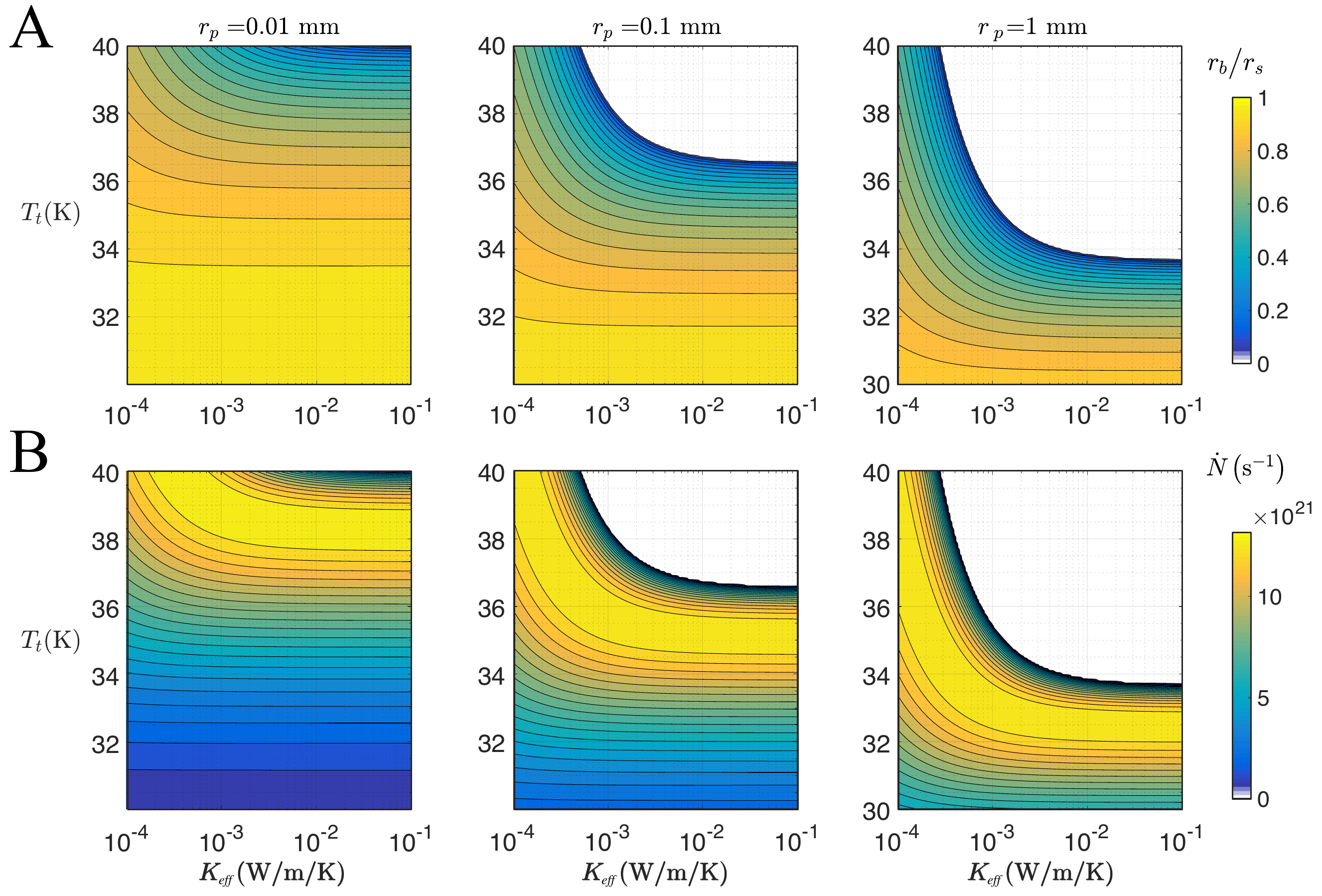}
\caption{Evolution of the CO sublimation front depth and CO gas production rate \REVNEW{for an Arrokoth-sized object as a function of the orbitally averaged temperature at $r_t$ ($T_t$), the conductivity ($K$) and pore radius ($r_p$).} 
\REV{(A) The relative depth of the sublimation front ($\zeta$) after 4.55 Gyr for a range of assumed pore radii, initial temperatures, and conductivities (see Table \ref{values_table}). The ratio $r_b/r_s=1$ represents an Arrokoth that has not undergone sublimation (at t=0), while smaller values (blues) indicate greater sublimation; 
(B) The associated gas production rate ($\dot N$) after 4.55 Gyr, which is dependent on the volume of CO ice remaining within Arrokoth (panel A).}}
\label{depth_production}
\end{figure}

Based on brightness temperature measurements by \textit{New Horizons}' REX instrument and best fits to Arrokoth's thermal inertia \citep{bird_detection_2022,umurhan_near-surface_2022}, and estimates of Arrokoth's bulk porosity \citep{keane_geophysical_2022,mckinnon_solar_2020}, we derive a range of effective thermal conductivities $10^{-4} - 10^{-2} {\rm W/m/K}$ for Arrokoth. This range is similar to cometary $K_{{\rm eff}}$ values \citep{groussin_thermal_2019} and those assumed in other studies \citep[e.g.,][]{bouziani_cometary_2022}. \REV{The smaller end of the assumed $K_{{\rm eff}}$ range follows the finding that TNOs have small thermal inertias \citep{Lellouch_etal_2013}}.
However, some measurements of comets have reported conductivities as high as $10^{-1} {\rm W/m/K}$ \citep{groussin_thermal_2019}, which we adopt as our upper limit for completeness. These larger conductivities have only a minor impact on the retention of CO within Arrokoth-like objects (Figure \ref{depth_production}). Objects like Arrokoth typically have a small seasonal skin depth ($r_t$) that ranges from a few to tens of meters \citep{groussin_thermal_2019}. Based on estimates for Arrokoth's thermal skin depth of $<30$ meters \citep{umurhan_near-surface_2022}, we assume that the ratio of $r_t$ to $r_s$, denoted as $\psi$, is approximately 0.998.

\begin{figure}[h!]
\noindent\includegraphics[width=\textwidth]{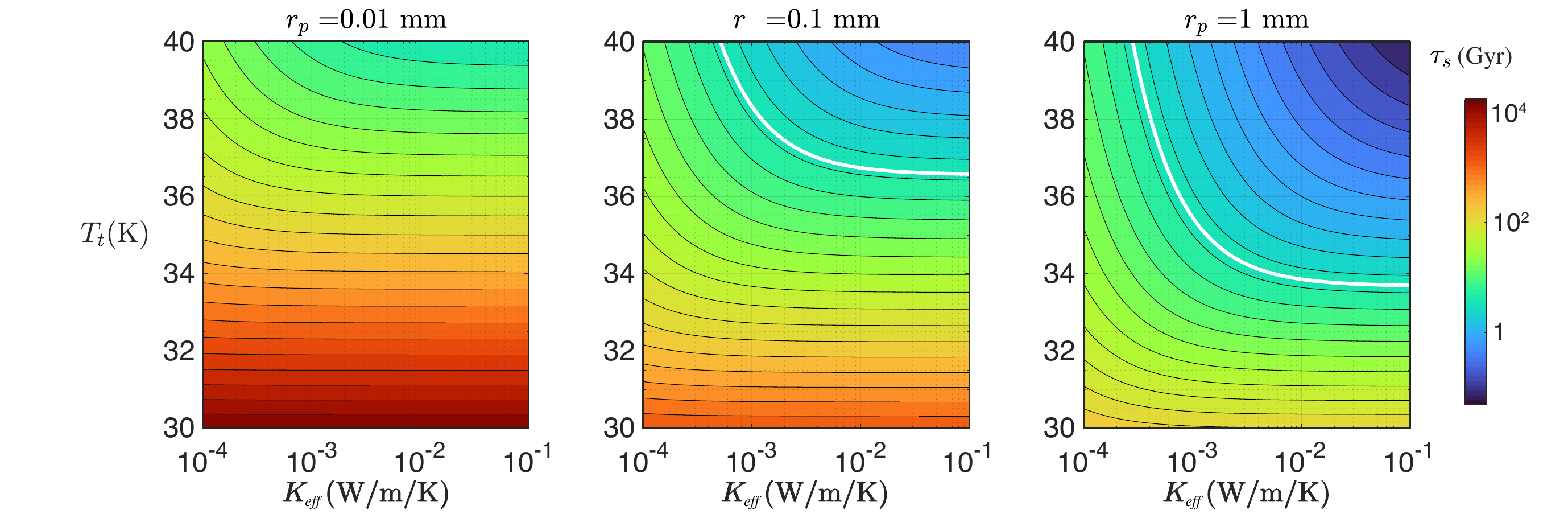}
\caption{Lifetime of CO within a porous, Arrokoth-sized object as a function of the orbitally averaged temperature at $r_t$ ($T_t$), the conductivity ($K$) and pore radius ($r_p$). Estimates of Arrokoth's orbitally averaged temperature range from 30$-$40 K \citep{umurhan_near-surface_2022}, with the most probable temperature near 34 K. For most scenarios, CO is retained within Arrokoth over the lifetime of the solar system (marked by a solid white line).}
\label{lifetime}
\end{figure}

\begin{figure}[h!]
\noindent\includegraphics[width=\textwidth]{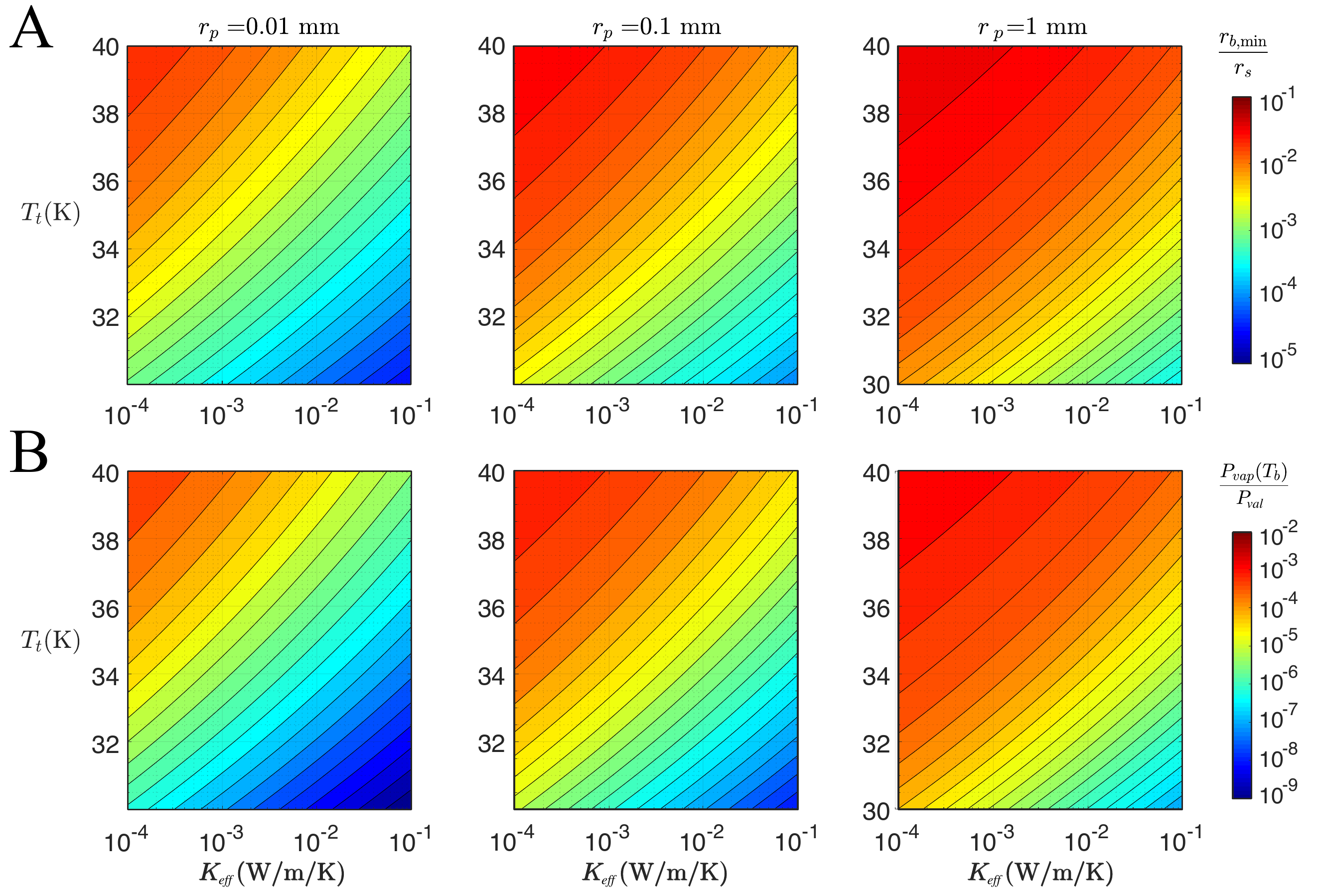}
\caption{\REV{Assessing the validity bounds of the quasi-steady evolution theory, which requires $\tau_t\ll\tau_{fr}$. 
(A) Values of $r_b \le r_{b,{{\rm min}}}$ at which the quasi-static theory breaks down. 
(B) The ratio $P_{{\rm vap}}(T_b)/P_{{\rm val}}$ for $r_b/r_s = 1/\varphi$, where $P_{{\rm val}} = P_{\varphi}$. Broadly, the condition $P_{{\rm vap}}(T_b)/P_{{\rm val}} \ll 1$ means that $\tau_t\ll\tau_{fr}$.  According to the trends shown in row B, the condition for validity are violated as $\Keff<<10^{-4}$, in which case  $\tau_t\gg\tau_{fr}$. In that event, the timescales reported in the previous figure should be considered as lower bounds.
We find that, quite generally, for the input parameters we have considered, both $r_{b,{{\rm min}}}/r_s \ll 0.1$ and $P_{{\rm vap}}(T_b)/P_{{\rm val}} \ll 1$.} \REVNEW{Variables and figure layouts are as in Figures \ref{depth_production} and \ref{lifetime}}}
\label{validity_figure}
\end{figure}

The range of possible pore sizes, which we assume to be proportional to the amorphous \water ice aggregate grain sizes, is still uncertain. In-situ measurements of dust grains conducted by \textit{Rosetta} at 67P range from 0.01$-$1 mm \citep{merouane_dust_2016}. Theoretical estimates provided by \cite{umurhan_near-surface_2022} align with these measurements. Therefore, we set $r_p$ to range from 0.01 to 1 mm (Table \ref{values_table}). Larger grains would not be consistent with the geophysical measurements of Arrokoth \citep{umurhan_near-surface_2022} and may violate our validity criterion (Section \ref{timescales_analysis} \& Figure \ref{validity_figure}). On the other hand, smaller grains will result in retention timescales longer than the age of the solar system for any reasonable temperature $T_t$ (see Figure \ref{lifetime}). 

Importantly, \cite{grundy_vaporpressures_2023} present novel laboratory measurements of CO vapor pressure that are significantly lower, by a factor of 5$\times$, compared to those collected in \cite{fray_sublimation_2009} and utilized in previous research analogous to ours \citep{bouziani_cometary_2022,lisse_origin_2021,Parhi_Prialnik_2023}. Hence, we utilize an approximate Arrhenius form for $P_{vap}$ that aligns with our temperature range of interest, based on the updated CO vapor pressure laboratory measurements \citep{grundy_vaporpressures_2023}. The form is given by
\beq
P_{vap}(T) = P\left(T_{{\rm{ref}}}\right)\exp\left(\frac{T_a}{T_{{\rm ref}}}-\frac{T_a}{T}\right),
\label{Pvap_Fit_CO}
\eeq
where $T_{{\rm{ref}}} = 30{\rm K}, P(T_{{\rm{ref}}}) \approx 6\times10^{-5} {\rm Pa},$ and an activation temperature of $T_a \approx 982$K. Notably, for the highest temperatures considered ($T=40$K), the updated vapor pressure value is found to be $P_{vap}(T=40 {\rm K}) \approx 0.22 {\rm Pa}$. 

We then estimate the total mass loss rate from Arrokoth as
\beq
\dot N = \dot \Sigma A_b = \dot \Sigma A_s \zeta^2,
\eeq
where $A_b$ represents the total area of the sublimating surface, which we assume to be equal to Arrokoth's total surface area (Table \ref{values_table}) reduced by a factor of $r_b^2/r_s^2$. Within our adopted temperature range, we operate in the particle diffusion limit (i.e., $\tilde P \gg P_{vap}(T_t)$), and we can simplify the expression for $\dot \Sigma$ given in Eq. (\ref{lebovsky}) (see Section \ref{sec:appendix}). This simplification leads to the total particle loss rate given by
\beq
\dot N(\zeta) \approx A_s \frac{3\bar c r_p}{8m v_k r_s} \cdot
\frac{\zeta}{1-\zeta} P_{vap}(T_b),
\eeq
where we evaluate $\zeta(t)$ using the solution provided in Eq. (\ref{implicit_zeta_solution}). We explore three different $r_p$ values and examine solutions for $\zeta$ within plausible ranges of $K$ and $T_t$ after 4.55 Gyr (Figure \ref{depth_production}A). We keep in mind the widely accepted timescale of 4.568 Gya for CAI emplacement \citep[][and references therein]{Dunham_etal_2022}.

Our analysis shows that for $r_p =$0.1mm \citep[or smaller;][]{bouziani_cometary_2022}, all primitive CO in Arrokoth can survive up to at least $T_t \approx 38$K, and can persist for the lifetime of the solar system (Figure \ref{depth_production}A/Figure \ref{lifetime}), within the plausible ranges of $\Keff$ and $T_t$ considered. \REV{Lower conductivities lead to even longer timescales.} 

This finding contradicts recent studies \citep{lisse_origin_2021,Parhi_Prialnik_2023} (Section \ref{Section4}), but should not be surprising, given the low conductivity, \REV{our assumed} lack of internal heat sources, and small pore radii in Arrokoth. Such conditions lead to a positive feedback loop: as temperature falls deeper inside the object, the rate of gas transport decreases, which, in turn, limits further increases in interior temperatures. If Arrokoth is relatively insulating, with an effective thermal conductivity of $\Keff \lessapprox 5 \times 10^{-3}$ Wm$^{-1}$K$^{-1}$, any initial CO inventory \REVNEW{(both gas within pores and ice at greater depths)} will be preserved almost indefinitely (Figure \ref{depth_production}A/Figure \ref{lifetime}). This is due to the slow movement of the sublimated gas, which effectively stays near vapor pressure equilibrium with ice at the front. Only for the largest pore sizes and highest conductivities does it become difficult to retain \REV{any} CO to the present day (Figure \ref{lifetime}), particularly when $r_p = 1$ mm and $T_t \gtrapprox 34$K (Figure \ref{depth_production}A). This is because gas more readily streams out of a matrix with such large pore sizes, interrupting the favorable feedback mechanism that arises for more reasonable pore radii. Nevertheless, CO \REVNEW{(both gas and ice)} can still be retained in the deeper regions of such objects for the entirety of the solar system's existence.

For all three values of $r_p$ considered, the maximum rate of CO molecules produced via sublimation after 4.55 Gyr ($\dot N \le 2\times 10^{22} {\rm s}^{-1}$) is also well below the \textit{New Horizons} upper limit $\dot N_{{\rm max}}$. We observe that as $T_t$ is varied and all other quantities are held fixed, the following trends emerge for $\dot N$ after 4.55 Gyr (Figure \ref{depth_production}B): (1) because the intrinsic sublimation rate is high for higher values of $T_t$, the CO sublimation front advances deeper into the interior, resulting in an effectively reduced surface area and a correspondingly low value of $\dot N$. (2) As $T_t$ is steadily reduced, the \REVNEW{front will not advance as significant of a distance into the interior}, and so the emitting surface area ($A_b = \zeta^2 A_s$) is larger, and $\dot N$ increases. (3) When $T_t$ is further lowered, the sublimation rate from the emitting front reduces substantially due to the Arrhenius dependence of $P_{{\rm vap}}(T_t)$. Despite the fact that the total emitting surface area is much larger, because the sublimation front has barely advanced, the net $\dot N$ decreases.

When examining the solutions generated for our input parameters, it is important to verify that they fall within the bounds of validity for our quasi-static theory, as summarized in Eq. (\ref{small_tildeP_validity}) for the $\tilde P\ll P_{vap}(T_b)$ limit that is relevant here. 
Figure \ref{validity_figure}B shows that the ratio $P_{{\rm vap}}(T_b)/P_{{\rm val}} \ll 1$ when estimated at $r_b/r_s = 1/\varphi \approx 0.618$.  We have argued earlier that assessing  $P_{{\rm vap}}(T_b)/P_{{\rm val}}$ is satisfactory to estimate general validity, and we therefore find that the criterion for validity is met across the entire parameter range considered. 
Figure \ref{validity_figure}A displays the limiting value $r_{b,{{\rm min}}}$, which represents the value of $r_b$ below which the quasi-static theory fails to hold (Section \ref{arrokoth_validity_section}). Since we are operating in the $P_{{\rm vap}}(T_b)/P_{\varphi} \ll 1$ regime, $r_{b,{{\rm min}}}$ can be well approximated by the expression given in Eq. (\ref{rb_min_definition}). Our results reveal that only for the highest values of $T_t$ and lowest values of $\Keff$ does $r_{b,{{\rm min}}}/r_s$ approach 0.05. Overall, throughout our analysis, $r_{b,{{\rm min}}}/r_s$ consistently stays well below 0.01. Only in the extreme scenario of the hottest and most conductive end-member, does our theory break down within the final few percent radii of the object near the object's core, which the sublimation front very rarely reaches anyway (Figure \ref{depth_production}).
\REV{Judging from the trends inferred upon examining Figure \ref{validity_figure}B, only when $\Keff \rightarrow 10^{-5}$ Wm$^{-1}$K$^{-1}$\citep[as in the extreme bookend case examined in][]{Kral_etal_2021} do we expect our lifetime predictions as possibly failing. Such low conductivites would mean $\tau_t\ge\tau_{fr}$, in which case our lifetime estimates should be interpreted as lower bounds instead. Nevertheless, our calculated timescales at these low conductivities often remain longer than the age of the solar system, as in \cite{Kral_etal_2021}.} 

\REV{Finally, it is important to note that we assume that the sublimation of CO is dictated by CO-CO bonding energies expressed via $T_a$. We assume that planetesimals like Arrokoth are comprised of mm-scale aggregates of $\mu$m-sized refractory grains (whether they be silicate based or water, or a mixture) that are either coated with CO or are intermingled with similar $\mu$m-sized CO grains. This picture has its basis in various global solar nebula models of grain growth in the outer solar system near and out beyond the CO ice-line \citep[like those discussed in][]{Birnstiel_etal_2012,Estrada_etal_2016,Estrada_Cuzzi_2022}.  
However if the component $\mu$m-sized grains are comprised instead of nm-scale intimately mixed CO-\water complexes, then the CO sublimation rates will change significantly due to differences in CO-\water binding energies that arise from mono-layer scale adsorption of CO molecules upon \water-ice substrates \citep[e.g.,][and several others since]{Kouchi_1990, Sandford_Allamandola_1990}. For example, \cite{Sandford_Allamandola_1990} report experimental results showing that CO-\water binding energies correspond to $T_a \approx 1740 \pm 50$K, which is not only far higher compared to that of CO-CO bonding ($T_a \approx 940$K), but would also lead to CO retention timescales significantly longer than those calculated here. We leave treating this possibility to a more comprehensive follow-up examination. Nevertheless, we consider the existence of nm-scale intimately mixed CO-\water grains to be unlikely because it would mean that both CO and \water molecules condensed out of the solar nebular gas at the same location at the same time within the protoplanetary disk, despite different condensation temperatures. Such a scenario seems unlikely, as it would require highly variable spatio-temporal temperature structures within the solar nebula.\footnote{\REV{We note that the experiments of \cite{Sandford_Allamandola_1990} involved the creation of intimately mixed CO-\water ice mixtures by spraying pre-mixed CO-\water gas onto the substrate inside their cryochamber.  It is fair to argue that this experimental setup would represent an astrophysical scenario in which both molecular species are simultaneously condensing.}} 
In all likelihood, $\mu$m scale homogeneous grains formed at different disk locations and only subsequently would turbulent disk transport result in grain species mixing across the disk \citep[e.g., like framed in][and others]{Estrada_etal_2016}.}

\section{Discussion} \label{Section4}
We demonstrated that primitive CO (and other hypervolatile) gas and ice reservoirs may exist within the interiors of Arrokoth and similar KBOs \REVNEW{(10's of km in diameter or less)}. Such reservoirs could have significant implications for how CO and amorphous \water ice interacted within the protoplanetary disk, and whether a gradual leakage of CO from Arrokoth's interior may alter Arrokoth's surface in the present day \citep[e.g.,][]{grundy_color_2020}. Confirmation of our theoretical predictions could be obtained by observing CO around other KBOs using upcoming space- and ground-based telescopes. Such observations would serve as a crucial validation of our work and allow for more detailed calculations of the internal evolution of these objects.

\subsection{Model Limitations}
Throughout our analysis, we have been mindful of the limitations of our analytical treatment, which is only valid under certain conditions due to the cold temperatures within Arrokoth's interior (Figure \ref{arrokoth_figure}B). Notably, our assumptions may not hold for Jupiter Family Comets in the inner solar system. However, in the remote outer solar system, the assumptions we used allowed us to simplify a complex non-linear system involving a series of partial differential equations into a single ordinary differential equation. We are not introducing any new physics compared to detailed models like NIMBUS \citep{davidsson_thermophysical_2021} or those of \cite{Prialnik_etal_2004}, instead we offer a simpler approach to obtain similar solutions -- appropriate in the cold limit -- thereby avoiding the need for computationally expensive models. In fact, we anticipate that our analytical solutions could be used to validate such models under the extreme thermophysical conditions that we are dealing with here.

Further investigations of KBOs could explore the influence of internal heat sources in their interiors where heating timescales remain long, such as heating from short- and long-lived radiogenic nuclides or the crystallization of amorphous \water ice, on our predictions \citep{malamud_are_2022, Parhi_Prialnik_2023}. In our application for a KBO like Arrokoth, we assumed that active radionuclides like $^{26}$Al were absent at the time of its assembly because KBOs may have formed quite possibly after 4 Myr after solar system formation \citep{bierson_using_2019}, which is several half-lives of $^{26}$Al. Similarly, Arrokoth is far too distant for amorphous \water ice to crystallize. KBOs, and similar objects, also have complex shapes and variations in their spin-orbit evolution that could cause asymmetries and dichotomies in the depth to the sublimation front. Arrokoth itself has a complex, bi-lobate shape (Figure \ref{arrokoth_figure}), and it is likely that such effects would be common on other KBOs \citep{jutzi_formation_2017,showalter_statistical_2021}. Investigating the combined effects of such factors, though unlikely to significantly change the overall conclusions outlined in Section \ref{Section3}, would permit a more general application to the broader family of KBOs.

\subsection{Comparison to Recent Similar Studies}
\REV{Our work differs from other recent studies \citep{Parhi_Prialnik_2023,lisse_origin_2021,lisse_predicted_2022,Kral_etal_2021} that investigated similar questions. \cite{Kral_etal_2021} explored the idea of retaining volatile ices within KBOs for billions of years. They employed an analytical approach similar to ours and arrived at a similar prediction: volatiles like CO might persist within KBOs until the present time. However, our conclusions are based on different considerations. According to \cite{Kral_etal_2021}, the rate of volatile loss is primarily governed by the time it takes for solar radiation absorbed at the surface to penetrate and drive sublimation deep within the interior. They further explain that $\tau_t \gg \tau_{fr}$ is generally true for KBOs owing to their very small thermal inertias, for which they require very low values of $\Keff = \order{10^{-5}}$Wm$^{-1}$K$^{-1}$, corresponding to the very lowest values of thermal inertia for TNOs reported in \cite{Lellouch_etal_2013}. These values of the conductivity may not be representative of the bulk interior of such objects \footnote{\REV{For example, comet 67P is thought to have a stratified interior with substantially higher effective values of thermal inertia \citep{groussin_thermal_2019}.}}. 
With such low thermal conductivies, the thermal relaxation timescale may indeed be comparable to or even greater than $\tau_{fr}$, leading to the dominance of $\tau_t$ in controlling the loss rate. Under these conditions, our lifetime estimates serve as conservative lower bounds, or equivalently, our predicted loss rates can be viewed as upper bounds.
}  

\REV{However, we disagree with the suggestion that  $\tau_t \gg \tau_{fr}$ for generic KBOs, as \cite{Kral_etal_2021} claim. Our findings indicate that, for the range of reasonable $r_p$ values we have explored, along with $\Keff = \order{10^{-4}}$ Wm$^{-1}$K$^{-1}$, the situation is quite the opposite: $\tau_t$ is considerably smaller than $\tau_{fr}$. In fact, when we examine the estimates for $\tau_t$ and $\tau_{fr}$ provided in Appendix A of \cite{Kral_etal_2021}, we observe that their calculations suggest gas diffusion times ($\tau_d\sim 10^4$ yr) that are even slower than their estimated front sublimation timescales ($\sim 10^3$ yr). The discrepancy in the two timescale estimates stems from the physical assumptions underlying their treatment of $\tau_{fr},$ which assumes that sublimation occurs in proportion to the enhanced surface area of an interstitial volatile ice block with effective pore spacing $r_p$ \citep[also see][]{Prialnik_etal_2004}. However, this treatment overlooks how a gaseous subsurface atmosphere with pressure $P$ — which itself steadily diffuses outward at a comparatively slow rate — acts to reduce the net sublimation from a volatile ice block. 

Specifically, their estimate for $\tau_{fr}$ fails to consider that the rate of volatile sublimation is also controlled by $P_{vap} - P,$ as expressed in the right-hand side of Eq. (\ref{lebovsky_label}). \REVNEW{They estimate a rapid CO sublimation timescale, $\tau_{{\rm subl-CO}}$, according to analogous formulae based on the free-streaming \water sublimation expression $\tau_{{\rm subl-H2O}}$ found in Section 3.9 of \citet{Prialnik_etal_2004}. But this timescale measures how rapidly a sublimating volatile ice layer {\textit{equilibrates to the ambient volatile gas}}. For the parameter range in which our theoretical framework is valid, this {\textit{relaxation rate}} is always rapid. But the corresponding volatile loss rate from the body is instead controlled by the rate at which the gas leaks out toward the surface. As a result, their sublimation model misattributes the slow liberation of gas molecules from the body to the extended time required for the thermal energy driving sublimation to reach the depths where the sublimating ice is located, instead of it being due to the very long time it takes for the sublimated gas molecules to diffuse toward the surface. 

This is the essence of the picture we have arrived at in our study: that volatile gases like CO in the interiors of primitive bodies like Arrokoth are likely in nearly exact vapor pressure equilibrium with their source volatile ices and their inexorable loss from the body is controlled by outward diffusion.}}

 Counter to our findings, \cite{ Parhi_Prialnik_2023} predict that CO should be severely depleted in under 100 Myr for spherical 5km KBOs at Arrokoth's heliocentric distance, and in just under 400 Myr for $\sim 10$km spherical KBOs (see their Table 3). \REV{Both their study and ours adopt characteristic $\Keff$ values that are typical of comets, which may be higher for TNOs \citep{Lellouch_etal_2013}.} \REVNEW{Both studies also adopt similar temperatures for the upper boundary condition.} However, there are three primary differences. First, the vapor pressure profiles that \cite{Parhi_Prialnik_2023} use in their analysis predict values of $P_{vap}$ for given $T$ that are $\approx 5-10 \times$ larger than the updated ones we use \citep[i.e., those of ][]{grundy_vaporpressures_2023}. \REVNEW{Second, it is difficult to precisely compare results, as key parameter values are not noted in their work (e.g., pore sizes).} Finally, they note that $^{26}$Al heating is a key agent in driving off CO and, also, is responsible for keeping the interior temperatures very low throughout the CO depletion process ($\sim 25$K). Thus, in the framework of \cite{Parhi_Prialnik_2023}, not only is $\tau_{fr} \approx \tau_t$, which is outside the validity regime of our analysis, but their timescales fall far to the upper right in any of the calculations shown in Figure \ref{lifetime}. We anticipate that our results could be recovered by their models if similar assumptions were made to those we make above. 

Our conclusions also differ from those of \cite{lisse_origin_2021, lisse_predicted_2022} who predict the total loss of hypervolatiles in KBOs like Arrokoth. In these studies, it is assumed that the mass loss rate is given by the free-streaming flux of sublimated vapor straight to the surface without any ambient gas pressure control on sublimation at the front that typify atmospheres in near-vapor pressure equilibrium, like Mars and Pluto. In other words, it is assumed that the sublimated gas does not diffuse through the volatile depleted refractory matrix above the sublimation front but, instead, flows straight out to the surface. Below, we provide a brief description as to how our work differs from \cite{lisse_origin_2021, lisse_predicted_2022}, with extra derivations detailed in Section \ref{app:2_lisse}.

Indeed, a free-stream state is a distinct possibility if the ambient gas pressures in such objects are larger than the grain-grain bonding strengths ($\sigma_g \sim $kPa) or grain-aggregate to grain-aggregate bonding strengths ($\sigma_{agg}\sim 1$Pa, also see discussion in Appendix A), thereby leading to the disintegration of the refractory subsurface porous matrix. However, we find that such high vapor pressures are not possible for bodies like Arrokoth at its current heliocentric distance of 45 A.U., and assuming KBOs (or some fraction of them) were formed well after radionuclides like $^{26}$Al have long burned out.

Nevertheless assuming a free-stream flux is like setting the ambient gas pressure ($P_b$) at the front to zero in Eq. (\ref{lebovsky_label}) and Eqs. (\ref{lebovsky}-\ref{dotrb_formal}), which is akin to identifying $\order{\tau_{fr}} = \order{\tau_{{\rm subl-CO}}}$ (see above).
Thus, the corresponding free-streaming mass flux, $\dot\Sigma_{free}$, would then be
\beq
\dot\Sigma_{free} \approx \frac{v_k}{c_s^2} P_{vap}(T).
\label{Sigma_free}
\eeq
In our treatment, the low-density gas flows through a porous medium with a corresponding mass-loss rate $\dot\Sigma_F$. In Section \ref{app:2_lisse} we develop a simple calculation that estimates $\dot\Sigma_F$ based on Fick's Law, which we show captures the spirit of the calculation done in our study in the Kn$> 1$ molecular flow regime limit. A direct comparison between the estimate for $\dot\Sigma_F$ found in Eq. (\ref{Sigma_F_appx}) and Eq. (\ref{Sigma_free}) immediately shows that $\dot\Sigma_F = (r_p/r_s) \cdot\Sigma_{free}$. While in both approaches the gas sublimation rate is controlled by $P_{vap}(T)$, allowing gas to free-stream toward the surface underestimates the volatile lifetime in a body by many orders of magnitude.

In general, we expect that a free-stream description will only be valid when the interior temperatures rise high enough that the internal gas pressures exceed $\sigma_{agg}$. In those interior regions where $P>\sigma$, the dynamics of the system may also resemble those of terrestrial fluidized beds, where particle-gas momentum exchange physics begin to play a role, thereby requiring a more careful treatment. 

\section{Acknowledgments}
The authors thank Will Grundy for providing us preliminary data regarding updated CO vapor pressure measurements. We also thank Bjorn Davidsson, Jason Soderblom, Jordan Steckloff for helpful guidance and reviewing earlier forms of this work. We also thank Stephanie Menten and an anonymous reviewer for helpful comments that improved this paper. This research was supported by the Heising-Simons Foundation (51 Pegasi b Fellowship to S.P.D.B.) and the \textit{New Horizons} Kuiper Belt Extended Mission I (support to O.M.U.). During the preparation of this work the author(s) used OpenAI's ChatGPT for proofreading purposes. After using this tool, the author(s) reviewed and edited the content as needed and take(s) full responsibility for the content of the publication.
\par
\begin{figure}[h!]
\noindent\includegraphics[width=0.9\textwidth]{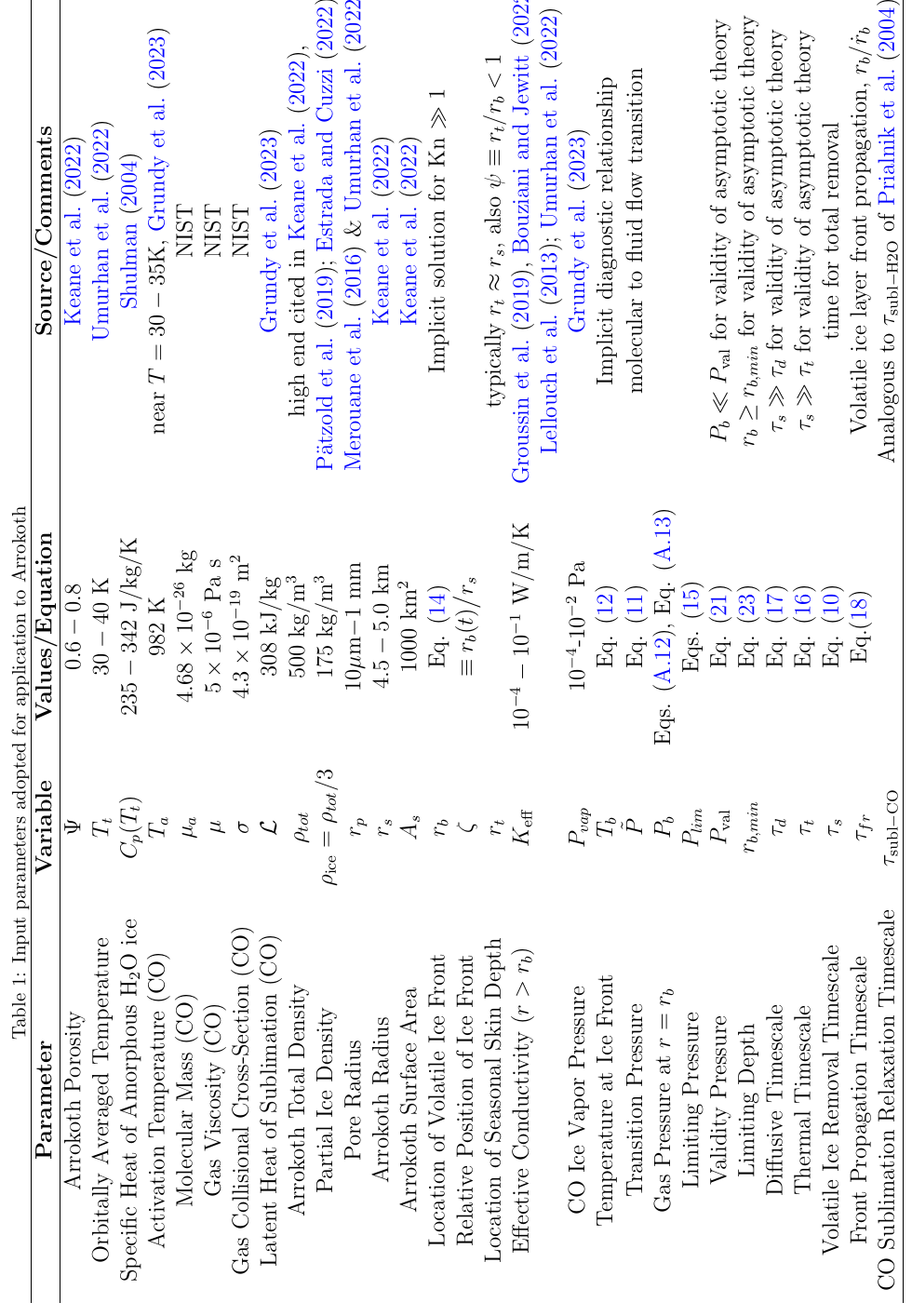}
\caption{Table}
\label{table}
\end{figure}

\appendix 

\section{Additional Mathematical Derivations \& Quantification of Assumptions} \label{sec:appendix}
We base our spherically symmetric model on the premise that sublimation \REVNEW{(or more precisely, the total CO depletion)} processes occur over much longer timescales compared to both thermal and dynamical readjustment times. Thus, we assume that there exists a \REV{time-variable radial location $r_b(t)$} beneath the surface at $r=r_s$ that represents the sublimating surface or `front' as described in the main text. The front gradually moves downward with a time rate of change $\dot{r_b}$ such that the corresponding timescales, $r_b/\dot{r_b}$ ($\tau_{fr}$), are significantly longer than the corresponding thermal wave propagation times ($\tau_t$) and dynamical readjustment times ($\tau_d$), as discussed in Section \ref{timescales_analysis}.

Our model adopts an interior structure where CO ice and amorphous \water ice intermix to form a porous and permeable matrix. The size of the void spaces within this matrix is $r_p$, and the porosity of Arrokoth ($\Psi_{tot}$) is uniform, ignoring any pore closure with depth.

We further assume that all the relevant physics occur for $r\ge r_b$ and that the ice layer at $r<r_b$ instantaneously adjusts to the sublimation dynamics at $r=r_b$. These assumptions may not hold if there are rapid changes to the system. However, we assume Arrokoth's orbital elements or energetic forcings do not change too quickly. Such matters could be addressed in follow-up studies that employ a more comprehensive approach.

In this quasi-static evolutionary framework, the time derivatives in the fluid equations of motion are neglected, resulting in a series of steady state equations for the system
\beqa
\frac{1}{r^2}\frac{\partial}{\partial r} \rho r^2 u
&=& \dot\Sigma \delta(r-r_b) \label{mass_conservation} \\
0&=& -k_a\frac{\partial P}{\partial r} - \mu u \label{radial_momentum} \\
0 &=& \frac{1}{r^2}\frac{\partial}{\partial r} 
K_{{\rm eff}} r^2 \frac{\partial T}{\partial r}
+ \dot Q,
\label{heat_equation}
\eeqa
where Eqs. (\ref{mass_conservation}-\ref{heat_equation}) represent: \REV{(1) mass conservation for the gaseous component with density $\rho(r,t)$, radial gas velocity $u(r,t)$, and a mass source $\dot \Sigma_b$ at the time variable location $r=r_b$; (2) momentum conservation in a generalized Darcy flow with gas pressure $P(r,t)$, dynamical molecular viscosity $\mu$, and generalized matrix permeability $k_a$; and (3) heat balance with temperature $T(r,t)$, effective conductivity $K_{{\rm eff}}$, and a heat sink term $\dot Q$, which we represent in terms of a Dirac delta function loss term localized at $r=r_b$}
\beq
\dot Q = -\dot{{\cal Q}} \delta(r-r_b),
\label{dotQ_def}
\eeq
to represent the consumption of energy via sublimation. We provide a detailed description of each equation, as well as the boundary conditions adopted, below. 

We make the additional assumption that there exists an enrichment of CO ice at a greater depth $r=r_b$, extending all the way to the center of Arrokoth at $r=0$ (Figure \ref{model_cartoon}). As a result, we assume that for all locations above this CO ice front at $r=r_b$, the porosity ($\Psi$ as referenced in the main text) is uniform but higher than its interior (i.e., $\Psi > \Psi_{\rm{tot}}$). {\REV{The temperature at the surface of the CO ice layer is designated as $T_b(t)=T(r=r_b,t)$}}. All of the physics we consider occur either at $r=r_b$ or above. We expect cold trapping to rapidly close any fluid pathways a short distance below $r=r_b$ and assume that no sublimated CO gas penetrates deeper than $r=r_b$. Hence, by construction, we assume flow only occurs radially outwards from $r_b$.

We proceed by developing solutions to Eqs. (\ref{mass_conservation}-\ref{heat_equation}) sequentially, beginning with the statement of mass conservation. The CO ice surface at $r=r_b$ sublimates with a rate 
$\dot \Sigma$ (in units kg$\cdot$m$^{-2}\cdot$s$^{-1}$) given by the standard formulations of \cite{lebofsky_stability_1975}
\beq
\dot \Sigma = \Big({P_{{\rm vap}}(T_b) - P_b}
\Big)\frac{v_K}{c_s^2},
\label{SigmabEqn}
\eeq
$c_s$ represents the sound speed with $c_s^2 = R_gT/\mu_a$, where $R_g = 8310$ J/kg/K is the gas constant and $\mu_a$ is the averaged atomic weight of the gas molecules (for CO $\mu=28$). The mean-averaged kinetic speed of the sublimated gas is $v_K$ and given by $\sqrt{8/\pi} c_s$. \REV{$P_b$ denotes the ambient gas pressure at $r=r_b$, i.e., $P_b(t) = P(r_b,t)$, and $P_{{\rm vap}}$ is the saturation vapor pressure of CO at temperature $T=T_b$}. In the case of Eq. (\ref{mass_conservation}), integration can be immediately performed to derive a constant mass loss rate $\dot M_0$ for $r\ge r_b$
\beq
\frac{\dot M_0}{4\pi} = \dot \Sigma r_b^2 
= \rho u r^2, \longrightarrow
\rho u = \frac{r_b^2}{r^2}\dot \Sigma .
\label{eqn:continuity}
\eeq
An immediate consequence of the mass conservation equation is an explicit expression for the mass flux $\rho u$. To simplify the analytical treatment, we assume that $c_s$ and $v_K$ are constant values that depend only on the temperature $T_t$ at the base of the seasonal thermal skin depth of Arrokoth (further discussed below in relation to Eq. (\ref{heat_equation})). For small KBOs like Arrokoth, the error introduced by this approximation is negligible and does not significantly affect the results presented in the main text. However, future studies could more thoroughly investigate this minor effect.

Let us now focus on Eq. (\ref{radial_momentum}), which describes the steady-state gas flow through the CO ice-depleted matrix permeability (i.e., $r>r_b$) using a generalized Darcy-Knudsen model. To account for the transition from low to high Knudsen number flows, we adopt the empirically motivated Klinkenberg formulation
\beq
k_a = k_\infty\big(1+4\bar c \rm{Kn}\big) ,
\label{Klinkenberg_Form}
\eeq
where
$
{\rm Kn} \equiv {\lambda}\big/{r_p},
$
which has been shown to be effective in similar studies \citep[e.g.,][]{ziarani_knudsens_2012}. Here, Kn represents the Knudsen number defined as the ratio of the gas mean free path, $\lambda$, to the pore radius, $r_p$, with $\lambda$ being defined in terms of the gas number density $n$ and a molecule's cross-section $\sigma$. The empirical constant $\bar c$ is typically around 1 and is assumed to be so in our study.

As suggested by \cite[][and references therein]{bouziani_cometary_2022}, we set the Darcy `liquid' permeability limit as $k_\infty = r_p^2/32$, where $r_p$ represents the pore radius. It is worth noting that the transition from `liquid' to `diffusive' flow usually occurs when Kn$>0.1$, but we use the more generalized form since Kn can vary depending on the pore sizes and vapor pressures under consideration. Nonetheless, previous studies related to comets, such as \cite{Espinasse_etal_1991}, have employed similar approaches that bridge both regimes.

Additionally, it is expected that the amorphous \water ice grains comprising Arrokoth's solid matrix possess adequate adhesive bonding such that the pressure of the sublimated CO gas is insufficient to displace them. This is true so long as $P\ll \sigma_p$, where $\sigma_p$ represents the bonding strength between pairs of individual grains or between pairs of grain-aggregates. Theoretical predictions on the strength of grain-grain contacts indicate that such materials are substantially stronger than the pressures of the sublimating gases \citep[e.g., JKR theory][]{JKR_1971}. Nonetheless, we also rely on strength assessments based on \textit{Rosetta} and \textit{Philae} measurements of comet 67P's surface materials \citep{biele_mechanical_2022} and years of extensive laboratory studies \citep[summarized in Figure 7 of][]{groussin_thermal_2019}. It has been found that grain-grain bonding for individual grains with radii in the few $\mu m$ scales is $\sigma_p \approx 1$kPa \citep[e.g.,][]{Gundlach_etal_2018}, while for mm-scale aggregates of these same $\mu$m scale grains, it is weaker, at $\sigma_p \approx 1$Pa \citep[e.g.,][]{Blum_etal_2014}. Assuming the conservative value for $\sigma_p$ (i.e., we assume that the primary constituents of interior particles are grain aggregates), we observe that the interior gas pressures are at least 3 orders of magnitude less for the temperature range of interest here ($30-40$K) based on recent CO vapor pressure measurements by \cite{grundy_vaporpressures_2023}. We therefore consider the sublimated vapor to be incapable of moving grains around and assume that the solid matrix (and in turn pore radii) remain fixed through time. As the gas diffuses or flows towards the surface, it is valid to use the Darcy-Knudsen flow law, Eq. (\ref{radial_momentum}).

Combining the relationship for $\rho u$ in Eq. (\ref{eqn:continuity}) with the Klinkenberg form in Eq. (\ref{Klinkenberg_Form}) and the assumption that $c_s$ is independent of temperature $T$, we obtain Eq. (\ref{radial_momentum}). This equation allows us to develop a first-order differential equation for $P$ as a function of space
\beq
\left(P + \frac{1}{2} \tilde P\right)\frac{\partial P}{\partial r} = -\frac{r_b^2 v_K \mu}{r^2 k_\infty}
\Big({P_{{\rm vap}}(T_b) - P_b}\Big),
\label{eqn:Darcy_reexpressed_1}
\eeq
where we have defined a transition pressure
\beq
\tilde P \equiv \frac{8\bar c m c_s^2}{\sigma r_p} .
\eeq
We can integrate Eq. (\ref{eqn:Darcy_reexpressed_1}) once, while taking into account the boundary condition that the pressure at $r=r_b$ is equal to an unknown constant $P_b$. This leads us to an implicit \REV{relationship for the function $P(r)$ valid for $r>r_b$,}
\beq
P^2 + \tilde P P = 
\frac{2r_b v_K \mu}{k_\infty}
\Big(P_b - P_{{\rm vap}}(T_b)\Big)
\left(1-\frac{r_b}{r}\right)
+ P_b^2 + \tilde P P_b .
\label{P_relationship}
\eeq
\REV{This expression for $P(r)$ is found in terms of the unknown $P_b$,  whose value can be solved if we demand that the pressure is zero at $r=r_s$ -- i.e., by setting $P(r=r_s)=0$ in Eq. (\ref{P_relationship}).  This condition leads to a relationship that must be satisfied between $P_b$ and the other parameters of the system:}
\beq
P_{{\rm lim}}(r,r_b) \Big(P_b - P_{{\rm vap}}(T_b)\Big)
+ P_b^2 + \tilde P P_b = 0; \qquad
P_{{\rm lim}} \equiv
\frac{2r_s v_K \mu}{k_\infty}
\frac{r_b}{r_s}\left(1-\frac{r_b}{r_s}\right).
\label{Plim_def}
\eeq
\REVNEW{The quantity $P_{{\rm lim}}$ is a quantity with units of pressure, but its physical meaning remains unclear.}
Solving the above for $P_b$ leads to
\beq
2 P_b = P_{{\rm lim}} + \tilde P
+ 
\left[(P_{{\rm lim}} + \tilde P)^2
+ 4 P_{{\rm lim}} P_{{\rm vap}}(T_b)
\right]^{1/2}.
\label{Pb_general}
\eeq
In the event that $P_{{\rm lim}} \gg \left\{P_{{\rm vap}}(T_b),\tilde P\right\}$, a Taylor series analysis shows that 
\beq
P_b \approx 
P_{{\rm vap}}(T_b)\left[
1 - \frac{P_{{\rm vap}}(T_b)+\tilde P}{P_{{\rm lim}}}
\right].
\label{Pb_appx}
\eeq
As previously noted, for bodies such as Arrokoth that are thought to contain CO, $P_{{\rm lim}}= \order{10^7\rm  Pa}$, which is consistently higher than both $P_{{\rm vap}}(T_b)$ (for temperatures $T$ in the range of $30-40$ K) and $\tilde P$ which typically is $\order{10^2}$ Pa for pore radii between 0.01 mm and 1 mm.

We can also determine the transition from the diffusive to the fluid regime (i.e., Kn $\approx \order{0.1}$) based on the relative magnitudes of $\tilde P$ and $P_{{\rm vap}}(T_b)$. In particular, if $P_{{\rm vap}}(T_b) \gg \tilde P$, the flow is in the fluid regime, whereas if $P_{{\rm vap}}(T_b) \ll \tilde P$, the flow is in the diffusive regime. For Arrokoth, we find that we are always in the diffusive regime, which implies that gas flow rates are generally suppressed. Models like NIMBUS \citep{davidsson_thermophysical_2021}, implicitly work within the diffusive regime as well. 

Next, we proceed to solve the heat equation (Eq. \ref{heat_equation}) to obtain the temperature at $T_b$. We assume that the subsurface low-density gas is in thermodynamic equilibrium with the refractory amorphous \water ice matrix through which it diffuses towards the surface. Therefore, we seek solutions for the temperature of the static matrix structure. To do so, we must first present a model for the effective conductivity of the matrix, which we assume to have the general form derived in \cite{umurhan_near-surface_2022}, given by
\beq
\Keff = 
K_c + K_r,
\eeq
where $K_c$ is thermal conductivity of the solid amorphous \water ice matrix and $K_r$ is the radiative conductivity across pore spaces within the matrix. For the former we adopt
\beq
K_c = K_A(1-\Psi) h,
\eeq
where $K_A$ is the conductivity of amorphous \water ice and $h$ is the adhesive fractional contact area given by Johnson Kendall Roberts theory \citep{JKR_1971}. We estimate value of $h$ to range from 0.01 to 1, while $K_A$ is experimentally found to be approximately 0.01 W/m/K \citep[see discussion after Eq. (38) of][]{umurhan_near-surface_2022}. This yields an effective range of $10^{-4} {\rm W/m/K} < K_c < 10^{-2} {\rm W/m/K}$, consistent with values used in other studies (e.g., \cite{bouziani_cometary_2022}). The radiative conductivity is typically assumed to be
\beq
K_r = 8 \epsilon_{IR} \sigma_{B}r_p T^3 ,
\label{radiative_conductivity_app}
\eeq
where $\sigma_{B}$ is the Stefan-Boltzmann constant and $\epsilon_{IR}$ is the infrared emissivity, which is typically around 0.9. The radiative conductivity ($K_r$) is linearly dependent on the pore size $r_p$, as shown in Eq. (\ref{radiative_conductivity_app}). However, for our assumed values of $T$ and $r_p$ (Table \ref{values_table}), we find that $K_r$ is much smaller than $K_c$, so we neglect the $K_r$ dependence and assume that $K_{{\rm eff}} = K_c$, which we treat as a constant. We assume values of $K_{{\rm eff}}$ as large as $10^{-1} {\rm W/m/K}$ for completeness, as discussed in the main text. 

As in previous, analogous studies on comets such as \cite{Espinasse_etal_1991} and \cite{Orosei_etal_1995}, we neglect heat conduction through the gas under the expected rarefied and dilute conditions of Arrokoth's interior. Here, we provide a brief justification. Using elementary Chapman theory to estimate molecule-molecule energy transfer, we can estimate the thermal conduction in a gas as
\beq
K_{gas} \approx \rho C_p \lambda v_K = \frac{C_p m v_K}{\sigma},
\eeq
where $\lambda$ is the collisional mean-free path (as defined earlier). Based on the range of values reported by \cite{Shulman_2004} for the specific heat capacity $C_p$ in the relevant temperature range (see Table 1), the gas conductivity is expected to be on the order of $\order{10^{-3}}$ W/m/K. However, this estimate is only valid if $\lambda \ll r_p$. In the case where $\lambda \gg r_p$ (as is likely the case for small KBOs like Arrokoth), molecules collide more frequently against pore walls, and energy transfer between molecules rarely occurs. For example, we can estimate that at $T_b = 35$K, CO's $P_{{\rm vap}} \approx 10^{-4}$Pa, which gives $\lambda \approx mc_s^2/(\sigma P_{{\rm vap}}(T_b))$. Using this approximation, we find that $\lambda$ is on the order of 1m, which is much greater than the assumed range of $r_p$. Therefore, we can safely neglect heat conduction through the gas.

In our model of Arrokoth, which is assumed to be a sphere with a radius $r=r_s$, we consider that the seasonal skin depth occurs at $r=r_t$ (see Figure \ref{model_cartoon}). At this location, we set the temperature to be the average of the extreme high ($T=T_{{\rm max}}$) and low ($T=T_{{\rm min}}$) surface temperatures, such that $T_t\equiv T(r=r_t)=0.5(T_{{\rm max}}+T_{{\rm min}})$. This temperature serves as the boundary condition for the long-term evolution of the interior. We then integrate Eq. (\ref{heat_equation}) for $r>r_b$, and express the temperature solution $T(r)$ in terms of an unknown basal thermal flux $F_b$
\beq
T(r) = \frac{F_b}{\Keff}\frac{r_b^2}{r_t}
\left(1-\frac{r_t}{r}\right) + T_t, \qquad
{\rm for} \ \  r>r_b,
\label{Tr_solution}
\eeq
while beneath the sublimation front we have the constant solution
\beq
T(r) = T_b, \qquad
{\rm for} \ \  r\le r_b.
\label{Tr_below_solution}
\eeq
Indeed, the solution for $r\le r_b$ represents the only possible steady-state thermal solution within the interior that avoids a singularity at $r=0$ that, in turn, implies that the thermal flux approaches zero as one approaches the sublimation front from below, i.e.,
\beq
K_{{\rm eff}} \frac{\partial T}{\partial r}\bigg|_{r\rightarrow r_b^{-}} = 0.
\label{flux_beneath_rb}
\eeq
Once again, this can be understood as a result of the very long timescales for the front to move compared to the timescales for thermal adjustment of the medium. In other words, since $\tau_{fr}\gg\tau_t$, the region inside $r<r_b$ has had enough time to reach a steady state in which it has received all the thermal energy it can hold. Once this steady state is reached, the thermal flux immediately below the front becomes zero, indicating that no further thermal energy can propagate inside (detailed in Section \ref{timescales_analysis} of the main text).

The volumetric energy loss rate $\dot Q$ defined in Eq. (\ref{dotQ_def}) is represented by a Dirac delta function centered at $r=r_b$. Here, $\dot{\cal{Q}}$ is the energy consumption rate per unit area due to sublimation at $r=r_b$, expressed in terms of $\dot \Sigma$ at $r=r_b$
\beq
\dot{\cal{Q}} = {\cal L}\dot\Sigma ,
\eeq
where the enthalpy of sublimation for CO ice is denoted by ${\cal L}$.
In our study, we assume that the only available energy to facilitate sublimation comes from thermal conduction, as given in Eq. (\ref{Tb_bc}) in the main text. Generally in this kind of treatment, the source energy is defined as the difference between the incoming thermal flux (as $r\rightarrow r_b^{+}$) and the outgoing thermal flux (as $r\rightarrow r_b^{-}$) at the front
\beq
K_{{\rm eff}} \frac{\partial T}{\partial r}\bigg|_{r\rightarrow r_b^{+}} 
- \ \ \ 
K_{{\rm eff}} \frac{\partial T}{\partial r}\bigg|_{r\rightarrow r_b^{-}} 
= \dot {\cal Q}.
\eeq
In light of the fact that the thermal flux beneath the sublimation front is zero (see Eq. (\ref{flux_beneath_rb})), the aforementioned condition provides an explicit expression that relates the unknown basal thermal flux $F_b$ to the rate of sublimation losses
\beq 
F_b = {\cal L}
\frac{v_K}{c_s^2}\Big[P_{vap}(T_b) - P_b\Big].
\label{Fb_solution}
\eeq
This expression relies on knowledge of the temperature $T_b$. To obtain the temperature $T_b$, we can substitute the expression for $F_b$ from the previous equation into Eq. (\ref{Tr_solution}). After some algebraic manipulation and utilizing the asymptotic form in Eq. (\ref{Pb_appx}), we arrive at the following equation for $T_b$
\beq
\frac{{\cal L}v_K}{{K_{{\rm eff}} c_s^2}}
\left[P_{vap}(T_b) + \tilde P\right]
\frac{P_{vap}(T_b)}{{P_{{\rm lim}}}} {r_b}\left(1-\frac{r_b}{r_t}\right)
= T_t-T_b,
\label{Tbrelationship}
\eeq
which concludes the complete description of the physical model.

Let us now focus on deriving the evolution equation for $r_b$. We start with the expression for the mass flux at $r=r_b$ (Eq. \ref{SigmabEqn}) and use the approximate solution in Eq. (\ref{Pb_appx}). We assume that the mass flux $\dot \Sigma$ can be expressed in terms of $\dot r_b$ (i.e., the rate of change of $r_b$) and the partial mass density of the sublimating ice, denoted as $\rho_{{\rm ice}}$. Thus, we write
\REVNEWW{\beq
 \dot r_b \rho_{{\rm ice}} = 
-\frac{v_K}{c_s^2} \left[P_{vap}(T_b) + \tilde P\right]
\frac{P_{vap}(T_b)}{{P_{{\rm lim}}}},
\label{dotSigma}
\eeq}
\REVNEWW{where the minus sign appearing on the RHS of Eq. (\ref{dotSigma}) ensures that $r_b$ recedes due to mass flux $\dot \Sigma$.}
By replacing $P_{{\rm lim}}$ according to its definition found in Eq (\ref{Plim_def}) and re-arranging the resulting expressions reveals that
\beq
\left(\frac{r_b}{r_s}\right)\left(1-\frac{r_b}{r_s}\right)
\left(\frac{\dot r_b}{r_s}\right) = 
\frac{1}{6\tau_s(T_b)},
\label{dotrb}
\eeq
which is equivalent to Eq. (\ref{rate}). Thus, by combining Eq. (\ref{Tbrelationship}) and Eq. (\ref{dotrb}), we have fully specified the time evolution of $r_b$.
 
Lastly, let us consider the issue of $\rho_{\rm ice}$, as the average density of KBOs can exhibit a wide range of values. Comets, which we use as an analogy, are typically assumed to have a density of $\rho_{tot} \approx 500$ kg/m$^3$ \citep[e.g.,][]{patzold_nucleus_2019}. According to \cite{keane_geophysical_2022}, the mean density of Arrokoth falls within the range $155$kg/m$^3 < \rho_{\rm arrokoth} < 600$kg/m$^3$, which overlaps with typical cometary values. Based on theoretical predictions of the compositions of planetesimals in the outer solar system at the time of their formation \citep[e.g., from solar nebula particle growth models such as those reported in][and references therein]{Estrada_Cuzzi_2022}, it is generally believed that hypervolatiles make up about a third of the total mass budget of a planetesimal, divided equally among silicates, water, and hypervolatiles such as CO. Therefore, we assume $\rho_{\rm ice} = \rho_{tot}/3 \approx 175$kg/m$^3$.

\section{Simplified diffusive flux derivation} \label{app:2_lisse}
The central physical effect controlling the loss of gas from the interior is gas diffusion according to Fick's Law. In the diffusion flow limit appropriate for Arrokoth (i.e., Kn$\gg 0.1$), the flow out from the interior is controlled by the pore spacing $r_p$ and the mean molecule speed $v_k$. The corresponding mass flux $\dot\Sigma_F$ relates to the gas diffusion coefficient $D = \order{r_p v_k}$ according to
\beq
\dot\Sigma_F  = D\frac{d\rho}{dr}, 
\eeq
which is basically Fick's Law for gas diffusion through a porous medium. Assuming there are no mass sources other than a sublimating front located a depth $\Delta r$ beneath the surface, and if we write the difference of the gas densities between the surface and the sublimation front as $\Delta \rho$, we can estimate $\dot\Sigma_F$ by writing
\beq
\dot\Sigma_F \approx r_p v_k \frac{\Delta \rho}{\Delta r} \approx 
r_p v_k \frac{P_{vap}(\Delta r)}{c_s^2 \Delta r},
\label{Sigma_F_appx}
\eeq
where we have approximated $\Delta\rho$ by saying that it is equal to the vapor density of the ice at the front ($\rho_{vap}$, estimated to be $P_{vap}/c_s^2$) minus the gas pressure at the surface, the latter of which is assumed to be zero.  

Equating the mass loss rate to a front propagation rate through an ice with density $\rho_{ice}$, we can further write a simple differential equation for the rate of change of $\Delta r$
\beq
\rho_{ice}\frac{d \Delta r}{dt} = \dot\Sigma_F.
\eeq
On the assumption that $P_{vap}(\Delta r)$ is constant because the temperature $T$ with depth does not vary much, the above expression, together with the use of Eq. (\ref{Sigma_F_appx}), may be integrated to derive a lapse time $\Delta t$ for the front to reach a depth $\Delta r$, i.e., 
\beq
\Delta t = \frac{1}{2}\frac{c_s^2\Delta r^2}
{\rho_{ice}r_p v_k P_{vap}(T)}.
\eeq
Setting $\Delta r \rightarrow r_s$ functionally recovers the diffusive limit of our result in Eq. (\ref{timescale}) with only minor $\order 1$  differences.

\bibliography{Comet_Gas.bib}
\bibliographystyle{aasjournal}

\end{document}